\documentclass[a4paper,10pt]{article}
\usepackage[english]{babel}
\usepackage[utf8]{inputenc}
\usepackage[T1]{fontenc}
\usepackage{amsmath,amsfonts,amssymb,amscd,amsthm}
\usepackage{xspace, xcolor, xifthen, xstring}
\usepackage{zref-savepos}
\usepackage{stmaryrd}
\usepackage{booktabs}
\usepackage[a4paper,top=3.5cm,bottom=3cm,left=2.5cm,right=2.5cm,bindingoffset=5mm]{geometry}

\usepackage[noblocks, auth-sc, affil-it]{authblk}
\usepackage{indentfirst}
\usepackage{microtype}
\usepackage{textcomp}
\usepackage{cite}
\usepackage[font=footnotesize,labelfont={bf}]{caption}%opzioni didascalie figure e tabelle
\usepackage{graphicx}%inserisci figure
\usepackage{subfig}%inserisci figure allineate
\usepackage{float}%scrivendo [H] dopo \begin{table} o \includegraphics{} non lo muove da quella posizione 
\usepackage{comment}
\usepackage{hyphenat}
\usepackage{array,longtable}
\usepackage{mathtools}
\usepackage{dsfont}
\usepackage[pdfpagelabels, hidelinks]{hyperref}
\usepackage[medium]{titlesec}
\numberwithin{equation}{section}
\usepackage{cleveref}
\crefname{equation}{equation}{equations}

\def\a{\alpha}
\def\b{\beta}
\def\g{\gamma}
\def\e{\,\mathrm{e}\,}
\def\l{\lambda}
\def\m{\mu}
\def\n{\nu}

\def\beq{\begin{equation}}
\def\eeq{\end{equation}}
\newcommand{\id}{{\mathds{1}}}
\newcommand{\im}{\,\mathrm{i}\,}
\newcommand{\diff}{\mathrm{d}}
\newcommand{\cf}{{\mathcal{F}}}
\renewcommand{\Re}{{\mathfrak{R}\mathrm{e}\, }}
\renewcommand{\Im}{{\mathfrak{I}\mathrm{m}\, }}
\def\+{\dagger}
\newcommand{\lb}{\left(}
\newcommand{\rb}{\right)}
\newcommand{\com}[2]{\left[#1,#2\right]}

\title{\LARGE{\textbf{Generalised Calibrations in AdS backgrounds from Exceptional Sasaki-Einstein Structures}}}

\author[$\dagger$,1]{\small{Oscar de Felice}}
\author[$\star$,2]{\small{Jakob C. Geipel}}

	\affil[$\dagger$]{Sorbonne Universités UPMC Paris 06, CNRS, UMR 7589, Laboratoire de Physique Théorique et Hautes Énergies, F-75005, Paris, France\vspace{2ex}}
	\affil[$\star$]{Institut f\"ur Theoretische Physik, Leibniz Universit\"at Hannover, Appelstra{\ss}e 2, 30167 Hannover, Germany}

\date{}
\renewenvironment{abstract}{\begin{center}\begin{minipage}{0.85\textwidth}\rule{\textwidth}{0.04em}\\[0.4em]\small\textbf{\abstractname~|}}{\par\rule{\textwidth}{0.04em}\end{minipage}\end{center}}
\begin{document}

\maketitle

%%%%%%%%%%%%%%%%%%%%%%%%%%%%%%
\begin{abstract}
	We study generalised calibrations described by Exceptional Sasaki-Einstein structures on AdS backgrounds in type IIB and M-theory, placing the focus on the generalised Reeb vectors.
	The inequalities for the energy bound are derived by decomposing a $\kappa$-symmetry condition and equivalently, bispinors in calibration conditions from existing literature.
	We explain how the closure of the calibration forms is related to the integrability conditions of the Exceptional Sasaki-Einstein structure, in particular for AdS space-filling or point-like branes.
	Doing so, we show that the form parts of the twisted vector structure in M-theory provides generalised calibrations, reproducing the result conjectured in a recent paper.
	The IIB case yields analogous results. 
	Finally, we discuss briefly the extension of the conjecture to hypermultiplet structures and other brane configurations. 
\end{abstract}
%%%%%%%%%%%%%%%%%%%%%%%%%%%%%%
%%%%%%%%%%%%%%%%%%%%%%%email addresses%%%%%%%%
{\let\thefootnote\relax\footnote{\hspace{-4ex}\textsuperscript{1}\emph{e-mail}: \href{mailto:odefelice@lpthe.jussieu.fr}{odefelice@lpthe.jussieu.fr}}}
{\let\thefootnote\relax\footnote{\hspace{-4ex}\textsuperscript{2}\emph{e-mail}: \href{mailto:jakob.geipel@itp.uni-hannover.de}{jakob.geipel@itp.uni-hannover.de}}}
%%%%%%%%%%%%%%%%%%%%%%%%%%%%%%%
\tableofcontents

\section{Introduction}

The interest for compactifications with non-trivial fluxes prompted the study of formulations of $10$ and $11$-dimensional supergravity where the gauge transformations of the higher-rank $p$-forms are treated on the same footing as diffeomorphisms. 
For instance, Generalised Complex Geometry (GCG) as defined by Hitchin and Gualtieri~\cite{Hitchin:2002ea, Gualtieri:2003dx}, and also referred to as $O(d, d)$ \emph{generalised geometry}, provides a very elegant description of the NSNS sector of type II supergravity theories by treating diffeomorphisms and gauge transformations of the NS two-form $B$ as elements of the $O(d, d)$ structure group of a generalised tangent bundle, where $d$ is the dimension of the compactification manifold.

The inclusion of non-vanishing RR fields requires a further extension of the generalised tangent bundle that now has structure group in the U-duality group. 
This leads to the notion of $E_{d(d)} \times \mathbb{R}^{+}$ or Exceptional Generalised Geometry~\cite{Hull:2007zu, Pacheco:2008ps, Coimbra:2011ky, Coimbra:2011nw, Coimbra:2012af}. 

A common approach to study supergravity backgrounds preserving a given amount of supersymmetry is the use of G-structures: the background admits a set of globally defined tensors that reduce the structure group. 
The supersymmetry conditions are reformulated as differential conditions on these geometrical structures. 
In generalised geometry, the same approach leads to the notion of generalised structures, which are now sum of forms (or polyvectors) of different degree. 
Again the supersymmetry equations can be expressed as differential conditions on these generalised structures. 
For instance, in Generalised Complex Geometry, the supersymmetry conditions become differential equations on a pair of polyforms that are pure spinors on the generalised tangent bundle~\cite{Grana:2004bg, Grana:2005sn}. 
A similar analysis in Exceptional Geometry led to the notion of \emph{Exceptional Calabi-Yau spaces (ECY)} for compactifications to Minkowski spaces~\cite{Ashmore:2015joa}. 

G-structures also appear naturally in defining calibration forms on the compactification manifolds. 
A $p$-form $\phi$ on a $d$-dimensional manifold $M$ ($d> p$) is a \emph{calibration} if and only if it is closed, \emph{i.e.} $\diff \phi =0$, and its pull-back to any tangent $p$-plane $\mathcal{S}$ satisfies the inequality
	\beq
	\label{eq:def_cal}
		P_{\mathcal{S}}[\phi] \leq \mathrm{vol}_{\mathcal{S}}\, ,
	\eeq
where $\mathrm{vol}_{\mathcal{S}}$ is the volume form on the plane $\mathcal{S}$ induced from the metric on $M$~\cite{Cal_Geo}. 
The ordering relation in~\eqref{eq:def_cal} has to be read as $P_{\mathcal{S}}[\phi] = \a \mathrm{vol}_{\mathcal{S}}$, with $\a \in \mathbb{R}^+$ and $\a \leq 1$.
For the supersymmetric backgrounds relevant in string and M-theory the calibration forms can be written as bilinears in the supersymmetry Killing spinors. 
For instance, on Calabi-Yau manifolds there are two types of calibration forms, which correspond to products of the K\"ahler form and to the real part of the holomorphic form on the Calabi-Yau. 

A $p$-dimensional submanifold $\Sigma_p$ is called \emph{calibrated} if it saturates the condition~\eqref{eq:def_cal} at each point: $P_{\Sigma_p} [\phi] = \mathrm{vol}_{\Sigma_p}$. 
One can show that a calibrated submanifold minimises the volume in its homology class. 
Indeed, given another submanifold $\Sigma'$, such that $\Sigma - \Sigma' =\partial B$ is the boundary of a $p+1$-dimensional manifold $B$, one has (see e.g.~\cite{Cal_Geo})
	\beq
		\mathrm{Vol}\lb \Sigma' \rb = \int_{\Sigma'} \mathrm{vol}_{\Sigma'} \geq \int_{\Sigma} P_{\Sigma}[ \phi] + \int_{\partial B} P_{\partial B}[\phi] 
			= \int_{\Sigma} \mathrm{vol}_{\Sigma} + \int_{B} \diff P_B[ \phi] = \mathrm{Vol} \lb \Sigma\rb\, , 
	\eeq
where we used the definition of calibration form, Stokes' theorem and the fact that $\Sigma$ saturates the inequality~\eqref{eq:def_cal}.
For a nice review on these arguments, one can refer, for instance, to~\cite{Joyce:2001nm}. 

Calibrations are useful tools in string theory because they provide a classification of supersymmetric branes in a given background. 
In a purely geometric background (no fluxes) supersymmetric branes wrap calibrated submanifolds, so that they minimise their volume~\cite{Becker:1995kb, Becker:1996ay, Gibbons:1998hm, Gauntlett:1998vk}. 
The calibration form is constructed as a bilinear in the Killing spinors of the background geometry, and its closure follows from the Killing spinor equations of the background. 

In the more general case of a background with non-trivial fluxes supersymmetric branes are associated with \emph{generalised calibrations}. 
Since the branes couple with the background fluxes, they do not correspond to minimal volume submanifolds but to configurations that minimise the energy. 
As in the fluxless case the generalised calibration form is related to the Killing spinors of the background~\cite{Gutowski:1999iu, Gutowski:1999tu, Townsend:1999nf, Gauntlett:2001ur, Gauntlett:2002sc, Gauntlett:2003cy, MS03, Cascales:2004qp, HPS03, HPS04}. 
Also in these cases the calibration forms can be written as bilinears in the Killing spinors and the closure of the generalised calibration form can then be deduced from the Killing spinor equations of the supersymmetric background~\cite{Cascales:2004qp, Gutowski:1999tu, Martucci:2005ht, MS03}.

Due to the AdS/CFT correspondence, much effort has been spent in understanding the structure of AdS compactifications~\cite{Gauntlett:2004zh, Gauntlett:2005ww, GGPSW09_02}. 
For these spaces exceptional generalised geometry yields the notion of \emph{Exceptional Sasaki-Einstein structures (ESE)}~\cite{Ashmore:2016qvs}. 
The exceptional structure contains a generalised vector $K$ that generalises the Reeb vector field and the contact structure of usual Sasakian geometry. 
For this reason, it is believed to encode information on brane configurations and conformal dimensions of chiral operators, as the contact structure does in~\cite{Martelli:2006yb}. 
In particular, in~\cite{Ashmore:2016qvs} the form parts of the generalised vector $K$ were conjectured to describe generalised calibrations for brane configurations dual to barionic operators in the dual gauge theory.
The aim of this work is to prove this conjecture and to show that the vector structure is indeed associated to generalised calibrations.
More precisely, we study the conditions to have supersymmetric extended objects in $\mathcal{N}=2$ AdS backgrounds in terms of Exceptional Sasaki-Einstein structures.

In this article, we are interested in investigating the relation between the Exceptional Sasaki-Einstein structures and generalised brane calibrations in $\mathrm{AdS}_5 \times M_5$ backgrounds in type IIB and in $\mathrm{AdS}_5 \times M_6$ and $\mathrm{AdS}_4 \times M_7$ compactifications in M-theory.
We focus on the calibration forms associated to branes wrapping cycles in the internal manifolds and that are point-like in the AdS space. 
We show that for these configurations the general expression for the calibration forms -- constructed using $\kappa$-symmetry -- can be expressed in terms of the generalised Killing vector $K$ defining the Exceptional Sasaki-Einstein structure. Moreover, we derive that the closure of the calibration forms is given by the integrability (more precisely the $L_K$ condition) of the ESE structure. 
Our results proves the conjecture appeared in~\cite{Ashmore:2016qvs}, that the generalised Killing vector is a generalised calibration.
We also partially discuss other brane configurations that are calibrated by the vector $K$.

The analysis in this article is far from being complete. For instance we did not fully study the calibration forms for branes with world-volumes spanning different directions in the AdS space. 
These should be related to components of the hypermultiplet structures and their closure to the moment map conditions. 
It would also be interesting to perform a similar analysis for compactifications to Minkowski space where the relevant structures are Generalised Calabi-Yau's~\cite{Ashmore:2015joa}. We leave this analysis for future work.

The article is organised as follows. In~\cref{sec:def_exceptional_geometry} we briefly review the basic notions of exceptional generalised geometry necessary in the following discussions.
A more detailed presentation can be found, for instance, in~\cite{Ashmore:2015joa, Ashmore:2016qvs}. 
Sections~\ref{sec:Mcal} and~\ref{sec:IIBcal} are the core of the work. 
We discuss the expression of the calibration forms and we interpret them in terms of Exceptional Sasaki-Einstein structures for $\mathrm{AdS}_4$ and $\mathrm{AdS}_5$ compactifications in M-theory (\cref{sec:Mcal}) and $\mathrm{AdS}_5$ compactifications in type IIB (\cref{sec:IIBcal}). 

Conventions for Clifford algebras and bilinears of spinor notations as well as some technical remarks on the approaches in the main text are relegated to the appendices. 
%
%
%======================================================================
%======================================================================
%++++++++++++++++++++++++++++++++++++++++++++++++++++++++++++++++++++++
%++++++++++++++++++++++++++++++++++++++++++++++++++++++++++++++++++++++
%======================================================================
%======================================================================
%
\section{Exceptional Geometry for M-theory and type II} 
\label{sec:def_exceptional_geometry}
In this section we briefly summarise the basic features of the generalised geometries we need in the rest of the paper. 
We refer to~\cite{Ashmore:2015joa} and~\cite{Ashmore:2016qvs} for more details. 

We are interested in compactifications of M-theory and type IIB supergravity to four and five dimensions with $\mathcal{N}=2$ supersymmetry (or equivalently with eight supercharges). 
In all these cases, we take the space-time to be a warped product 
	\beq
	\label{warpmetr}
		\diff s^2 = e^{2 \Delta} \diff s^2(X_D) + \diff s^2(M_d) 
	\eeq
where $M_d$ is a $d$-dimensional compact manifold and $X_D$ is the external space-time with $D=(11-d)$ in M-theory and $D=(10 - d)$ in type IIB. The external manifold $X$ has Lorentzian signature and, for our purposes, it will be Minkowski or Anti de Sitter. $\Delta$ is a real function of the internal coordinates called \emph{warp factor}.

In exceptional geometry one considers an extended tangent bundle on $M$ whose structure group is the exceptional group $E_{d(d)}$ in M-theory or $E_{d+1(d+1)}$ in type II and then defines geometric structures on it.
These groups correspond to the U-duality groups of the toroidal compactifications of the theories, in our cases to four or five dimensions.
%%%
%%
%
\subsection{M-theory}
The relevant groups for compactifications to four and five dimensions are $E_{7(7)}$ and $E_{6(6)}$, respectively. 
In both cases the generalised tangent bundle is a vector bundle of the group $E_{d(d)}$. It is convenient to decompose the generalised tangent bundle in representations of the structure group $GL(d)$ of the manifold $M$,
	\beq
	\label{Mgtb}
		E \cong \tilde{E} \coloneqq T M \oplus \Lambda^2 T^* M \oplus \Lambda^5 T^* M \oplus ( T^* M \otimes \Lambda^7 T^* M) \, . 
	\eeq
The sections of $E$ are called \emph{generalised vectors} and can be written as 
	\beq
		V =v+\omega + \sigma + \tau \, ,
	\eeq
where $v \in TM$ is a vector, $\omega \in \Lambda^2 T^* M$ is a two-form, $ \sigma \in \Lambda^5 T^* M$ a five-form and $ \tau \in T^* M \otimes \Lambda^7 T^* M$ is the tensor product of a one-form and a seven-form.\\
One can also introduce generalised tensors as sections of bundles carrying different representations of $E_{d(d)}$. 
In particular, one defines the adjoint bundle as 
	\beq
	\label{Madj}
		\mathrm{ad} F \cong \mathbb{R}_\Delta \oplus ( T M \otimes T^* M) \oplus \Lambda^3 T^* M \oplus \Lambda^6 T^* M \oplus \Lambda^3 T M \oplus \Lambda^6 T M\, ,
	\eeq
with sections 
	\beq
		\mathcal{A} = l + r + A + \tilde A + \a + \tilde\a \, ,
	\eeq
where $l \in \mathbb{R}$ gives the shift of the warp factor $\Delta$, $r \in End(T M)$, $A \in \Lambda^3 T^*M$ is related to the three-form potential of M-theory, $\tilde A \in \Lambda^6T^*M$ to its dual, while $\a \in \Lambda^3 TM$ and $ \tilde\a \in \Lambda^6 TM$ are a three- and a six-vector.

The three- and six-form potentials $A$ and $\tilde A$ of M-theory are encoded as twists of the generalised tangent bundle. 
The twisted bundle $E$ has sections
	\beq
	\label{twistV}
		V = e^{A + \tilde{A}} \tilde{V}\, ,
	\eeq
where $\mathcal{A}=\tilde A + A$ is in the adjoint of $E_{d(d)}$ and its (exponentiated) action is defined in appendix~$E$ of~\cite{Ashmore:2015joa}. 
The choice of potentials in~\eqref{twistV} defines an isomorphism between the twisted and untwisted bundles $E$ and $\tilde{E}$\cite{Pacheco:2008ps}. 
This constrains the topology of the generalised tangent bundle and gives the correct patching (or gauge) transformations for the M-theory potentials~\cite{Coimbra:2011ky}.\\
Similarly, if $\tilde{R}$ is a section of $\mathrm{ad} F$, its twisting is given by,
	\beq
		R = e^{A + \tilde A}\ \tilde{R}\ e^{-A - \tilde A} \, ,
	\eeq 
which is compatible with the twisting of the fundamental representation in~\eqref{twistV}.
%%%
%%
%
\subsection{Type IIB theory}
\label{eggIIB}
	The relevant groups in type II are again $E_{7(7)}$ and $E_{6(6)}$, and the generalised tangent bundle is then a vector bundle of $E_{d+1 (d+1)}$~\cite{Coimbra:2012af, Ashmore:2015joa}.
	The decomposition of $E$ under $GL(d)$ reads 
		\beq
		\label{IIBgtb}
			\begin{split}
 				E \cong \tilde{E} & \coloneqq\ TM \oplus T^*M \oplus \Lambda^{\mathrm{odd}} T^*M \oplus \Lambda^5 T^*M \oplus \lb T^*M \otimes \Lambda^6 T^*M \rb\\ 
 					&\cong\ TM \oplus \lb S \otimes T^*M\rb \oplus\Lambda^3 T^*M \oplus \lb S \otimes \Lambda^5 T^* M\rb \oplus \lb T^*M \otimes \Lambda^6 T^*M\rb\, , 
			\end{split}
		\eeq
	where $\Lambda^{\mathrm{odd}} T^* M$ denotes odd forms on $M$.
		\footnote{%
			For IIA the generalised tangent bundle is the same with odd forms replaced by even ones. 
			Equivalently, one can derive the EGG relative to type IIA from the M-theory one, reducing on a circle all the relevant quantities.%
			}%
	The sections of the untwisted bundle are 
		\beq
			\tilde{V}= v + \tilde{\l}^i + \tilde{\rho} + \tilde{\sigma}^i + \tilde{\tau} \, .
		\eeq
	The index $i=1,2$ on the one- and five-forms denotes that they transform as doublets of the $\mathrm{SL}(2)$ S-duality group. 
	In this case the adjoint bundle is 
		\beq
		\label{IIBadj}
			\mathrm{ad} F \cong \mathbb{R}_\Delta \oplus \mathbb{R}_\phi \oplus \Lambda^2 T^* M \oplus \Lambda^2 T M \oplus \Lambda^6 T^* M \oplus
			 					\Lambda^6 T M \oplus \Lambda^{\mathrm{even}} T^* M \oplus \Lambda^{\mathrm{even}} T M \, ,
		\eeq
	where $\mathbb{R}_\phi$ and $\mathbb{R}_\Delta$ denote the shifts of the dilaton and warp factors respectively, the forms correspond to the RR and NSNS potentials, while the vectors are obtained raising indices with the metric. 
	The twisted tangent bundle again follows upon twisting with elements of the adjoint bundle,
		\beq
			\begin{aligned}
				V = \e^{B^i + C} \tilde{V} \, ,&\phantom{and} & i = 1,2 \, , 
			\end{aligned}
		\eeq
	where $B^i$ are the NS and RR two-form potentials and $C$ is the four-form potential. 
%%%
%%
%
\subsection{Generalised Lie derivative}
	An important object in the construction of the theory is the generalised Lie derivative or Dorfman derivative.
	It generates the infinitesimal generalised diffeomorphisms (diffeomorphisms plus gauge transformations) and it is defined in analogy with the ordinary Lie derivative $\mathcal{L}$~\cite{Pacheco:2008ps}. 
	Given two generalised vectors $V'$ and $V$ the generalised Lie derivative is
		\beq
			L_V V' = V \cdot \partial V' - \left(\partial \otimes_{\mathrm{adj}} V \right) \cdot V' \, ,
		\eeq
	where $\cdot$ denotes the adjoint action, and the term in brackets indicates the projection onto the adjoint bundle representation,
		\beq
			\begin{array}{rcc}
				\otimes_{\mathrm{adj}}:& E^* \times E \rightarrow \mathrm{adj}\, . & 
			\end{array}
		\eeq
	Decomposing in $GL(d)$ representations, the action of $L_V$ on a generalised vector $V'$ is
		\beq
			\begin{split}
				L_V V' & = \mathcal{L}_v v' + \left(\mathcal{L}_v \omega' - \iota_{v'} \diff \omega\right) +\left(\mathcal{L}_{v} \sigma' -\iota_{v^\prime}\diff \sigma - \omega^{\prime}\wedge \diff \omega\right) \\
						&\phantom{=} + \left(\mathcal{L}_{v} \tau^{\prime} - j \sigma^{\prime}\wedge \diff \omega - j \omega' \wedge\diff \sigma \right) \, 
			\end{split}
		\eeq
	for M-theory\footnote{%
			Given $\lambda \in \Lambda^{p}T^*M$ and $\mu \in \Lambda^{d-p+1}T^*M$, the ``$j$-operator'' is defined as 
				\begin{equation*}
					\left(j \lambda \wedge \mu\right)_{m,\,m_1\ldots m_d} \,=\, \frac{d!}{(p-1)!(d-p+1)!}\,\lambda_{m[m_1\ldots m_{p-1}}\mu_{m_p\ldots m_d]}\ , 
				\end{equation*}
		where $j \lambda \wedge \mu \in T^*M \otimes \Lambda^d T^*M$.%
		}%
	~\cite{Pacheco:2008ps}, while for type IIB it reads~\cite{Ashmore:2015joa},
		\beq
			\begin{split}
				L_V V' & = \mathcal{L}_v v' + \left(\mathcal{L}_v \lambda^{i\prime} - \iota_{v'} \diff \lambda^{i}\right) \\
					&\phantom{=} + \left(\mathcal{L}_{v} \rho' -\iota_{v^\prime}\diff \rho +\epsilon_{ij} \diff \lambda^{i}\wedge \lambda^{j\prime}\right) \\
					&\phantom{=} + \left(\mathcal{L}_{v} \sigma^{i\prime} -\iota_{v^\prime}\diff \sigma^i + \diff \rho \wedge \lambda^{i\prime} - \diff \lambda^{i} \wedge \rho^{\prime} \right) \\
					&\phantom{=} + \left(\mathcal{L}_{v} \tau^{\prime} - \epsilon_{ij} j \lambda^{i\prime}\wedge \diff \sigma^j + j \rho' \wedge\diff \rho + \epsilon_{ij} j \sigma^{i\prime} \wedge \diff \lambda^j \right) \, .
			\end{split}
		\eeq

	In practical computations it is often useful to work in terms of untwisted quantities. We then define a ``twisted'' generalised Lie derivative that acts on an untwisted tensor $\tilde{\a}$ as
		\beq
			\hat{L}_{\tilde{V}} \tilde{\a} = \e^{-\mathcal{A}} L_{\e^{\mathcal{A}} \tilde{V}} \left(\e^{\mathcal{A}} \tilde{\a} \e^{-\mathcal{A}}\right)\, ,
		\eeq
	where $\mathcal{A}$ is the element of the adjoint bundle used to twist the generalised tensor $\tilde{\a}$. The twisted Dorfman derivative can be written as 
		\beq
		\label{eq:twisted_untwisted_lie_der}
			\hat{L}_{\tilde V} \tilde{\a} = \mathcal{L}_{\tilde v} \tilde{\a} - \tilde{R}( \tilde{V}) \cdot \tilde{\a}\, ,
		\eeq
where the adjoint element $\tilde{R}$ is given in terms of the untwisted vector $\tilde{V}$ and the fluxes of the potentials used for the twisting~\cite{Ashmore:2015joa}
		\beq
		\label{eq:tensor_r}
			\tilde{R}(\tilde{V}) =
				\begin{cases}
					\diff \tilde{\omega} - \iota_{v} F\ + \diff \tilde{\sigma} - \iota_v \tilde{F} +\tilde{\omega} \wedge F & \mbox{M-theory} \, , \\[1mm]
 					\diff \tilde{\l^i} -\iota_v F^i + \diff \tilde{\rho} - \iota_v F - \epsilon_{ij} \tilde{\l}^i \wedge F^j + \diff \tilde{\sigma}^i +\tilde{\l}^i \wedge F - \tilde{\rho} \wedge F^i &\mbox{type IIB} \, . 
				\end{cases} 
		\eeq
	Here, the gauge invariant field strengths are
		\beq
		\label{fluxes}
			\begin{array}{cccc}
				F = \diff A\, , & \tilde{F} = \diff \tilde{A} - \tfrac{1}{2} A \wedge F &\phantom{and} &\mbox{M-theory} \\
				F^i = \diff B^i\, , & F = \diff C -\tfrac{1}{2}\epsilon_{ij}F^i \wedge B^j & \phantom{and}&\mbox{type IIB}
			\end{array}
		\eeq
	where one has in M-theory $\tilde{F} = \star_{11} F$, and in type IIB $F^1_3 = H$ and $F^2_3 = F_3$.
%%%
%%%%%%%
%%%%%%%%%%%%%%%%%%%%%%
%%%%%%%
%%%
\subsection{Exceptional structures}
\label{ExStr}
	As for ordinary G-structures, the existence of globally defined generalised tensors reduces the structure group of $E$ and defines generalised $G$-structures in exceptional geometry.
	
	As shown in~\cite{Ashmore:2015joa, Ashmore:2016qvs, Grana_Ntokos}, a supersymmetric background with eight supercharges is characterised by the existence of hyper- and vector-multiplet structures, defining the relative generalised $G$-structure.\\
	A \emph{hypermultiplet structure}~\cite{Ashmore:2015joa}, or \emph{H structure} for short, is a triplet of sections of the weighted adjoint bundle 
		\beq
			J_a \in \Gamma (\mathrm{ad} \tilde{F} \otimes (\det T^*M)^{1/2})\, , 
		\eeq
	such that
	\beq
		\begin{array}{ccccc}
			\com{J_a}{J_{b}} = 2 \kappa \epsilon_{abc} J_{c}& & \mbox{and} & & \mathrm{tr} \lb J_{a}J_{b}\rb =- \kappa^2 \delta_{ab}\, .
		\end{array}
	\eeq
	The triplet $J_{a}$ realizes a $\mathrm{Spin}^*(12) \subset E_{7(7)} \times \mathbb{R}^+$ structure and an $\mathrm{SU}^*(6) \subset E_{6(6)} \times \mathbb{R}^+$ for compactifications to four and five dimensions, respectively.
	
	A \emph{vector structure}, or \emph{V structure}, is given by a generalised vector
		\beq 
			K \in \Gamma \lb E\rb 
		\eeq
	that has positive norm 
		\beq
			\begin{array}{ccc}
				q(K) >0 & \mbox{or} & c(K)>0\, , 
			\end{array}
		\eeq
	where $q(K)$ denotes the $E_{7(7)}$ quartic and $c(K)$ the $E_{6(6)}$ cubic invariant. The generalised vector $K$ defines an $E_{6(2)}$ and $F_{4(4)}$ structure in $D=4$ and $D=5$, respectively.

	One can impose the following compatibility conditions on $J_a$ and $K$
		\beq
		\label{compHV}
			\begin{array}{ccc}
				J_{a} \cdot K=0 & \mbox{and}& \mathrm{tr}\lb J_{a}J_{b}\rb =
					\begin{cases}
						-2 \sqrt{q(K)} \delta_{ab}&\phantom{\mbox{for}} D=4\\
						-c(K) \delta_{ab}&\phantom{\mbox{for}} D=5 
					\end{cases}
			\end{array}
		\eeq
	The pair $(J_{a}, K)$ is then called an \emph{HV structure} and defines an $\mathrm{SU}(6) = \mathrm{Spin}^*(12) \cap E_{6(2)}$ structure and a $\mathrm{USp(6)} = \mathrm{SU}^*(6) \cap F_{4(4)} $ structure in $D=4$ and $D=5$, respectively (see~\cite{Ashmore:2015joa}). The explicit form of the $HV$ structure depends on the theory and the dimension of the compactification manifold. 
	For instance, the generalised vector $K$ is 
		\beq
		\label{genvecs}
			K = \left\{ \begin{array}{lcl} 
						\xi + \omega + \sigma + \tau & \phantom{\mbox{for}} & \mbox{M theory} \\[1mm]
						\xi + \lambda^i + \rho + \sigma^i + \tau & \phantom{\mbox{for}} & \mbox{type IIB}
					\end{array} \right.
		\eeq
	We will give the form of the $H$ structure in the following sections for the cases of interest for this work.\\

	As discussed in~\cite{Ashmore:2015joa, Ashmore:2016qvs}, the Killing spinor equations for backgrounds preserving $\mathcal{N}=2$ supersymmetry are equivalent to a set of integrability conditions for the $HV$ structure $(J_a, K)$,
		\begin{subequations}
			\begin{align} 
				\label{eq:moment_map}
				&& &\m_{a} (V) = \l_{a} \g(V) && \forall \, V \in \Gamma(E)\, , && \\
				\label{eq:LK1}
				&& &L_{K}K=0\, ,\\
				\label{eq:LK2}
				&& &L_K J_a = \epsilon_{a b c} \l_{b} J_{c} \, , && L_{\tilde K} J_{a} =0 \, , &&
			\end{align}
		\end{subequations}
	where the second condition in~\eqref{eq:LK2} only applies for $D=4$. The functions $\m_{a} (V)$ are a triplet of moment maps for the action of the generalised diffeomorphisms,
	\beq
		\mu_a(V) = - \frac{1}{2} \epsilon_{abc} \int_M \mathrm{tr}(J_b L_V J_c)\, .
	\eeq
	The constants $\lambda_a$ are zero for Minkowski backgrounds, while for AdS are related to the inverse of the AdS radius $|\l|=2m$ for $D=4$ and $|\l|=3m$ for $D=5$, where $|\l|^2=\l_1^2+\l_2^2+\l_3^2$. Finally, the function $\gamma$ is defined as
		\beq
			\begin{array}{llc}
				\gamma (V) = 2 \int_M q(K)^{-1/2} q(V,K,K,K) &{}& D=4\, , \\[1mm]
				\gamma (V) =\int_M c(V,K,K) & {}& D=5 \, .
			\end{array}
		\eeq

	In this paper we are interested in AdS compactifications. 
	The AdS backgrounds satisfying~\eqref{eq:moment_map}-\eqref{eq:LK2} are called \emph{Exceptional Sasaki-Einstein (ESE)}~\cite{Ashmore:2016qvs}. Examples of such structures are type IIB theory on $\mathrm{AdS}_5 \times M_5$ and M-theory on $\mathrm{AdS}_5 \times M_6$ and $\mathrm{AdS}_4 \times M_7$~\cite{Ashmore:2016qvs}, where they generalise the properties of the usual Sasaki-Einstein manifolds. 
 
	For \emph{ESE} structures, an important consequence of the supersymmetry conditions is that the generalised vector $K$ is a generalised Killing vector, that is
		\beq
			L_K \mathcal{G} = 0	\, ,
		\eeq
	where the generalised metric $\mathcal{G}$ contains the bosonic degrees of freedom of the supergravity theory~\cite{Coimbra:2011ky,Coimbra:2012af}. The generalised Killing vector condition for M-theory is equivalent to
		\beq
			\begin{array}{ccccc}
				\mathcal{L}_\xi g = 0\, , &\phantom{and}& \mathcal{L}_\xi A - \diff \omega = 0\, , &\phantom{and}& \mathcal{L}_\xi \tilde{A} -\diff \sigma + \tfrac{1}{2} \diff \omega \wedge A = 0\, ,
			\end{array}
		\eeq
while in type IIB one has
		\beq
			\begin{array}{lcl}
				\mathcal{L}_\xi g = 0\, , &\phantom{and}& \mathcal{L}_\xi C = \diff \rho - \tfrac{1}{2}\epsilon_{ij}\diff \lambda^i \wedge B^j \, , \\[1mm] 
				\mathcal{L}_\xi B^i - \diff \lambda^i = 0\, , &\phantom{and}& \mathcal{L}_\xi \tilde{B}^i = \diff \sigma^i + \tfrac{1}{2} \diff \lambda^i \wedge C - \tfrac{1}{2} \diff \rho \wedge B^i + \tfrac{1}{12}\epsilon_{kl} B^i \wedge B^k \wedge \diff \lambda^l \, .
			\end{array}
		\eeq
	%
	%Corresponding to the known gauge transformations on the potential fields in both M-theory and type IIB.

	The generalised Killing vector condition on $K$ means that the action of the generalised Lie derivative on the untwisted objects reduces to the usual one,
		\beq
		\label{dorflie}
			\hat{L}_{K} \cdot =\mathcal{L}_{\xi} \cdot \, , 
		\eeq
	where $\xi$ denotes the (necessarily non-vanishing) vector component of $K$. By virtue of~\eqref{eq:twisted_untwisted_lie_der} this is equivalent to the vanishing of the tensor $\tilde{R}$ in~\eqref{eq:tensor_r}. 
	We will refer to~\eqref{eq:moment_map} as \emph{moment map condition} while to~\eqref{eq:LK1},~\eqref{eq:LK2} and the vanishing of $\tilde{R}$ in~\eqref{eq:tensor_r} as $L_K$ \emph{condition}.
	$K$ is called \emph{generalised Reeb vector} because it naturally generalises the isometry described by the usual Reeb vector in Sasakian geometry. 
%
%%
%
%%%%%%%%%%%%%%
%%%%%%%%%%%%%%
%
%
%======================================================================
%======================================================================
%++++++++++++++++++++++++++++++++++++++++++++++++++++++++++++++++++++++
%+++++++++++++++++++++++++++ M-Theory Section +++++++++++++++++++++++++++++
%======================================================================
%======================================================================
%
%
%
\section{Generalised calibrations in M-theory}
\label{sec:Mcal}

The aim of this section is to study calibrations for supersymmetric brane configurations of M-theory on AdS backgrounds of the form~\eqref{warpmetr} in terms of exceptional geometry.
AdS calibrations have been thoroughly discussed in the literature~\cite{Gutowski:1999iu, Gutowski:1999tu, MS03, Koerber:2007jb} and led to the notion of generalised calibration. 
In this section we will interpret these calibrations in terms of the Exceptional Sasaki-Einstein structures describing the AdS background.

Supersymmetry static M-theory backgrounds have been studied in~\cite{GauntlettGeomKill}. 
A supersymmetric background admits a Majorana Killing spinor $\varepsilon$ satisfying,
		\beq
			\nabla_M \varepsilon + \frac{1}{288} \left[ \Gamma_M^{\phantom{M}NPQR} - 8 \delta_M^{\phantom{M}N} \Gamma^{PQR} \right] G_{NPQR}\ \varepsilon =0 \, ,
		\eeq
where $M,N, \ldots = 0,1, \dots, 10$, $G= \diff A $ is the four-form field strength and the Gamma matrices are the Clifford algebra elements in $11$ dimensions. 
The four-form $G$ and the metric $g$ satisfy the relative equations of motion
		\begin{align}
			R_{MN} - \dfrac{1}{12}\left( G_{MPQR}G_{N}^{\phantom{N}PQR} - \dfrac{1}{12}g_{MN} G^2 \right) &= 0 \, ,\\
			\diff \star G + \dfrac{1}{2} G \wedge G &= 0\, .
		\end{align}

	The Killing spinor can then be used to build one-, two- and five-forms
	\begin{subequations}
		\label{11df}
			\begin{align}
				\label{11d1f}
				\mathcal{K}_M &= \bar{\varepsilon} \Gamma_M \varepsilon \, , \\
				\label{11d2f}
				\omega_{M N} &= \bar{\varepsilon} \Gamma_{M N} \varepsilon \, , \\
				\label{11d5f}
				\Sigma_{MNPQR} & = \bar{\varepsilon} \Gamma_{M N PQ R} \varepsilon \, , 
			\end{align}
	\end{subequations}
and the supersymmetry conditions imply that
	\begin{subequations}
		\begin{align}
			& \diff \mathcal{K} = \tfrac{2}{3} \iota_\omega G + \tfrac{1}{3} \iota_\Sigma \star G \, , \\
			& \diff \omega = \iota_\mathcal{K} G \, , \\
			& \diff \Sigma = \iota_\mathcal{K} \star G - \omega \wedge G \, . 
		\end{align}
	\end{subequations}

Supersymmetry also implies that the vector $\hat{\mathcal{K}}^M$ dual to the one-form~\eqref{11d1f} is a Killing vector, \emph{i.e.}
		\beq
			\begin{array}{ccc}
				\mathcal{L}_{\hat{\mathcal{K}}} g = 0\, , & \phantom{and} & \mathcal{L}_{\hat{\mathcal{K}}} G = 0 \, . 
			\end{array}
		\eeq	
	The vector $\hat{\mathcal{K}}^M$ can be either null or time-like, and for the backgrounds of interest here it is time-like%
		\footnote{%
			In this case the forms $\mathcal{K}$, $\omega$ and $\Sigma$ define an $SU(5)$ structure in 11 dimensions.%
		}.%

	Now let us focus on AdS backgrounds,
		\beq
		\label{eq:metric_mtheory_ads}
			\diff s^2 = \e^{2\Delta}\diff s^2(\mathrm{AdS}) + \diff s^2(M)\, ,
		\eeq
	where $\Delta$ is a real function on $M$, the warp factor.
	
	As usual, to construct the generalised calibrations for M-branes we can use $\kappa$-symmetry. 
	A supersymmetric brane satisfies the bound
		\beq
		\label{kprojector}
			\hat{\Gamma} \varepsilon = \varepsilon \, ,
		\eeq
	where $\varepsilon$ is the background Killing spinor and the $\kappa$-symmetry operator $\hat \Gamma$ depends on the type of brane. For an $\mathrm{M}5$-brane this is defined as~\cite{MS03,Gabella:2012rc},
		\beq
		\label{KopMth}
			\hat{\Gamma} = \frac{1}{L_{DBI}}\Gamma_0 \left[\frac{1}{4}\Gamma^\a (\tilde{H}\lrcorner H)_\a + \frac{1}{2} \Gamma^{\a\b} \tilde{H}_{\a\b}+ \frac{1}{5!}\Gamma^{\a_1 \ldots \a_5} \epsilon_{\a_1 \ldots \a_5} \right]\, , 
		\eeq 
	where $H = d B + P[A]$ is the world-volume three-form, $\tilde{H}$ is its world-space dual~\cite{Pasti:1997gx,Pasti:1995tn,Bergshoeff:1998vx}%
		\footnote{%
			The field $\tilde{H}$ is defined in terms of an auxiliary scalar field $a$, which is needed to ensure the Lorentz covariance of the world-volume Lagrangian~\cite{Pasti:1997gx,Pasti:1995tn},
				\beq
					\tilde{H}_{\mu\nu} = \frac{1}{\sqrt{\vert\partial a \vert^2}} \left(\star H\right)_{\mu\nu\a}\partial^{\a} a (\sigma)\, .
				\eeq
			The scalar $a$ is subject to a gauge transformation and one usually fixes it by going to the \emph{temporal gauge}, \emph{i.e.} $a= \sigma^0 = t $. 
			This gauge fixing procedure breaks the Lorentz invariance $SO(1,5)$ down to $SO(5)$ and sets $\tilde{H}$ equal to the world-space dual of $H$.%
			}%
		and $L_{DBI}$ is the Dirac-Born-Infeld Lagrangian for the $\mathrm{M}5$ brane,
		\beq
		\label{DBIlagr}
			L_{DBI} = - \sqrt{-\det(P[g] + \tilde{H})}\, .
		\eeq 
	Per usual, $P[\bullet]$ denotes the pull-back on the $\mathrm{M}5$ world-volume and we defined
		\beq
			\Gamma_{\a_1 \ldots \a_s} = \Gamma_{M_1 \ldots M_s} \partial_{\a_1}X^{M_1} \ldots \partial_{\a_s}X^{M_s}\, .
		\eeq

	As discussed in~\cite{MS03, Gabella:2012rc}, the $\kappa$-symmetry condition~\eqref{kprojector} can be used to derive the following bound~\cite{Barwald:1999hx},
		\beq
		\label{M5ksym}
			\lVert \varepsilon \rVert^2 L_{DBI}\ \mathrm{vol}_5 \geq \left[ \frac{1}{2} P[ \iota_{\hat{\mathcal{K}}} H ] \wedge H + P[\omega] \wedge H + P[ \Sigma] \right]\, ,
		\eeq
	where $K$, $\Sigma$ and $\omega$ are defined in~\eqref{11df}.
	To satisfy the bound one has to take into account that the space is Anti-de Sitter. 
	As discussed in~\cite{Gabella:2012rc}, the norm $\varepsilon^\dagger \varepsilon$ depends on the AdS coordinates and the bound is saturated when the $\mathrm{M}5$ brane sits at the center of AdS.
	Explicitly, the metric of $\mathrm{AdS}_n$ in global coordinates can be written as,
		\beq
			\diff s^2 = R^2\left(-\cosh^2 \varrho\, \diff t^2 + \diff \varrho^2 + \sinh^2 \varrho\, \diff \Omega_{n-2} \right)\, ,
		\eeq
	and $\varepsilon^\dagger \varepsilon \propto \cosh \varrho$, thus, the bound~\eqref{M5ksym} can be saturated only for $\varrho = 0$, \emph{i.e.} in the center of AdS.

	Further, the bound~\eqref{M5ksym} can be used to derive a bound on the energy of an $\mathrm{M}5$ brane \cite{MS03,Gabella:2012rc}. The energy of the an $\mathrm{M}5$-brane is given by 
		\beq
		\label{energyM5}
			E_{\mathrm{M}5} =- \int_{\mathcal{S}} \diff^5 \sigma\ g(\hat{P},\hat{\mathcal{K}}) \, , 
		\eeq
	where $\mathcal{S}$ denotes the $5$-dimensional world-space of the brane, $\hat{P}_M$ is the conjugate momentum%
		\footnote{%
		To write this expression, we have chosen again the static gauge $X^M = ( t, \sigma^\a)$.%
		}~\cite{Bergshoeff:1998vx},
		\beq
			\begin{split}
				\hat{P}_M =& \frac{\partial L_{M 5}}{\partial (\partial_\tau X^M)} = P_M + \dfrac{1}{4} \dfrac{1}{5!} \epsilon^{\tau \a_1\ldots\a_5} H_{\a_1 \a_2\a_3}H_{\a_4\a_5\a_6} \partial^{\a_1} X_M \\
						&\phantom{= P_M + + } - \dfrac{\tau_5}{5!} \epsilon^{\tau \a_1\ldots\a_5}\left[ \iota_M \tilde{A} - \tfrac{1}{2} \iota_M A \wedge (A - 2 H)\right]_{ \a_1\ldots\a_5} \, ,
			\end{split}
		\eeq
	where $X^M$ are the embedding coordinates of the brane. 
	The quantity $g(\hat{P},\hat{\mathcal{K}}) = \hat{P}^M \hat{\mathcal{K}}^N g_{MN}$ can be interpreted as a Noether charge density of the symmetry generated by $\hat{\mathcal{K}}$~\cite{Martucci:2011dn}. 
	Then the inequality~\eqref{M5ksym} gives a bound on the energy of the brane,
		\beq
		\label{BPScon}
			E_{\mathrm{M}5} \geq E^{BPS}_{\mathrm{M}5}\, ,
		\eeq
	where
		\beq
			E^{BPS}_{\mathrm{M}5} = \int_{\mathcal{S}} P[\Sigma] + P[\iota_{\hat{\mathcal{K}}} \tilde{A}] + P[\omega] \wedge H + \tfrac{1}{2} P[\iota_{\hat{\mathcal{K}}} H] \wedge (A - 2H)\, .
		\eeq
	As shown in~\cite{Gabella:2012rc}, the form
		\beq
		\label{genkcal}
			\Phi_{\mathrm{M}5} = \Sigma + \iota_{\hat{\mathcal{K}}} \tilde{A}+\omega \wedge H + \tfrac{1}{2} \iota_{\hat{\mathcal{K}}} H \wedge (A - 2H)\, ,
		\eeq
	is a generalised calibration, namely is closed by supersymmetry and it minimises the energy in its homology class being a topological quantity~\cite{MS03}.

	The discussion for an $\mathrm{M}2$ brane works analogously, and the calibration form is 
		\beq
		\label{genkcalM2}
			\Phi_{\mathrm{M}2} = \omega + \iota_{\hat{\mathcal{K}}} H\, .
		\eeq

As final comment, we want just to point out that the construction above can also be derived by the supersymmetry algebra. The same calibration forms emerge in the supersymmetry algebra with the central extensions due to the presence of BPS extended objects, and one can prove their closure by using the Killing spinor equations~\cite{HPS03,Gutowski:1999tu, Cascales:2004qp}.

%%%
%%
%
\subsection{\texorpdfstring{Calibrations on $\mathrm{AdS}_5 \times M_6$}{Calibration on AdS5 x M6}}
Even if the formalism described in the previous section is completely general, in what follows we will focus on static M-branes in backgrounds of the type~\eqref{eq:metric_mtheory_ads} and we will show how the calibration forms~\eqref{genkcal} and~\eqref{genkcalM2} are naturally encoded in the generalised Sasaki-Einstein structure.

We consider first the case of compactifications to $5$-dimensional AdS spacetime. 
The supersymmetry conditions for backgrounds of this type are given in~\cite{Gauntlett:2004zh}, while the corresponding exceptional generalised geometry is presented in~\cite{Ashmore:2016qvs,Grana_Ntokos}, and we briefly review it below. 
We refer to these works also for notation and conventions, and we collect, for convenience, again the relevant conventions used here in~\cref{app:mtheoryconv}.

The metric takes the form
\beq
\label{eq:metric_mtheory_ads5}
	\diff s^2 = \e^{2\Delta}\diff s^2_{\mathrm{AdS}_5} + \diff s^2_{M_6}\, ,
\eeq
where we denote the inverse AdS radius as $m$. As shown in~\cite{Gauntlett:2004zh}, supersymmetry constrains the geometry of the six-dimensional internal manifold: $M_6$ has a local $SU(2)$ structure and topologically is a two-sphere bundle over a four-dimensional base%
		\footnote{%
			The four dimensional base can be either a K\"ahler-Einstein manifold with positive curvature or a product of two constant curvature Riemannian surfaces. The latter case in non-Einstein~\cite{Gauntlett:2004zh}.%
			}. 

There is a non-trivial four-form field strength $\mathcal{F}$ with non-zero components along the internal manifold $M_6$, 
	\beq
		F_{m_1 \ldots m_4} = \left(\mathcal{F}\right)_{m_1 \ldots m_4}\, ,
	\eeq
while the external components are set to zero, $\mathcal{F}_{\mu_1 \ldots \mu_4} = 0$.

The internal flux $F$ satisfies the equations of motion and the Bianchi identity
	\beq
		\begin{array}{ccccc}
			\diff F = 0\, , & & & & \diff (\e^\Delta \star_6 F) = 0\, , 
		\end{array}
	\eeq
with $\star_6$ the Hodge star on $M_6$, while the dual form $\tilde{F}_{m_1 \ldots m_7} = \left(\star_{11}\mathcal{F}\right)_{m_1 \ldots m_7}$ identically vanishes on the six-dimensional internal manifold $M_6$.

The Clifford algebra $\mathrm{Cliff}(1,10)$ decomposes in $\mathrm{Cliff}(1,4)$ and $\mathrm{Cliff}(0,6)$:
	\beq
	\label{eq:dec_gammas_m6}
		\begin{array}{ccc}
			\hat{\Gamma}^\mu = \e^{\Delta}\,\rho^\mu \otimes \gamma_7\, , & \phantom{and} &\hat{\Gamma}^{m+4} = \mathds{1}_4 \otimes \gamma^m\, ,
		\end{array}
	\eeq
with $\gamma_7 = - \im \gamma^1 \ldots \gamma^6$ the chiral operator in $6$ dimensions, and $\rho$ and $\gamma$ matrices satisfying
	\beq
	\label{eq:cliff}
		\begin{array}{ccc}
			\left\{\rho_\a, \rho_\b\right\} = 2 \eta_{\a\b} \mathds{1}\, , & \phantom{and} &\left\{\gamma_{a} , \gamma_b \right\} = 2 \delta_{ab} \mathds{1}\, ,
		\end{array}
	\eeq
in terms of the frame indices $\a,\b = 0,\ldots,4$ on $\mathrm{AdS}_5$ and $a,b = 1, \ldots, 6$ on $M_6$. We collect further conventions about spinors and Clifford algebras in~\cref{app:mtheoryconv}.

To have an $\mathcal{N}=2$ supersymmetric background we decompose the $11$-dimensional spinor as
	\beq
	\label{eq:decomp_mtheory_spinors}
		\varepsilon = \psi \otimes \chi + \psi^c \otimes \chi^c\, ,
	\eeq
where $\psi$ is an element of $\mathrm{Cliff}(1,4)$. 
Notice that, in order to have an AdS backgrounds, the internal spinor $\chi$ cannot be a chirality eigenstate~\cite{Gauntlett:2004zh}. 
Hence, it can be written as,
	\beq
	\label{chiredef}
		\chi = \sqrt{2} \left(\cos \a \chi_1 + \sin \a \chi_2^* \right)\, ,
	\eeq
where $\a$ is a parameter chosen to get the unit norm for the spinor, as in~\cite{Gauntlett:2004zh}.\\

The exceptional geometry for these backgrounds is given in~\cite{Ashmore:2016qvs,Grana_Ntokos}.
%\subsubsection{Exceptional Sasaki-Einstein structure}
The exceptional bundles to consider are again~\eqref{Mgtb} and~\eqref{Madj}. 
The vector structure $K\in E$ and the hypermultiplet structure $J_a \in \mathrm{ad}F$ can be expressed in terms of the $SU(2)$ structure of~\cite{Gauntlett:2004zh}. 

In this paper we are mostly concerned with the generalised Killing vector $K$. 
Its untwisted version is given by
	\beq
	\label{vecK}
		\begin{array}{cc}
			\tilde{K}=\xi -\e^{\Delta} Y' + \e^{\Delta} Z \equiv \xi + \tilde{\omega} + \tilde{\sigma} & \in \tilde{E} 
		\end{array}
	\eeq
with the vector $\xi$, the two-form $Y'$ and five-form $Z$ defined in terms of spinor bilinears as in~\cite{Ashmore:2016qvs}, 
	\begin{align}
		\label{vecbilM6}
		\xi &= -\im \left(\bar{\chi}_1+\chi_2^T\right)\gamma^{(1)}\left(\chi_1 - \chi_2^* \right)\, ,\\[1mm]
		\label{YbilM6}
		Y' &= -\im \left(\bar{\chi}_1+\chi_2^T\right)\gamma_{(2)}\left(\chi_1 - \chi_2^* \right)\, ,\\[1mm]
		\label{ZbilM6}
		Z &= -\im \left(\bar{\chi}_1+\chi_2^T\right) \gamma_{(5)} \left(\chi_1 - \chi_2^* \right)\, .
	\end{align}
The twisted version of the V structure is obtained by the (exponentiated) adjoint action
\beq
	K = \e^{A+\tilde{A}} \tilde{K}\, ,
\eeq
where $A$ is the three-form and $\tilde{A}$ the six-form potential of $M$-theory. Using the expressions for the commutator and the adjoint action from~\cite[App. E]{Ashmore:2015joa}, one obtains~\cite{Ashmore:2016qvs} 
	\beq
	\label{eq:twisted_K_M}
		K= \xi + \lb \iota_{\xi}A - \e^{\Delta}Y'\rb + \lb \e^{\Delta} Z - \e^{\Delta} A \wedge Y' + \tfrac{1}{2} \iota_{\xi} A \wedge A\rb\, .
	\eeq
As discussed above, the tensor $\tilde{R}$ must vanish for the generalised Lie derivative to reduce to the usual one, and this is equivalent to~\cite{Ashmore:2016qvs}
	\begin{align} \label{isom5}
		& &\diff \tilde{\omega} = \iota_{\xi} F\, , & & \diff \tilde{\sigma} = \iota_{\xi} \tilde{F} - \tilde{\omega} \wedge F\, . & &
	\end{align}
On $M_6$, this yields the differential conditions
	\begin{subequations}
	\label{eq:m-theory_structure_eqs}
		\begin{align}
			\diff \lb \e^{\Delta} Y'\rb &= - \iota_{\xi} F\, , \\[1mm]
			\diff \lb \e^{\Delta}Z\rb &= \e^{\Delta} Y' \wedge F\, ,
		\end{align}
	\end{subequations}
which we refer to as $L_K$ conditions in the language of exceptional generalised geometry.

Supersymmetry gives also the Killing vector condition
	\beq
		\mathcal{L}_\xi F = \mathcal{L}_\xi \Delta = \mathcal{L}_\xi g = 0\, . 
	\eeq

We can now discuss how the generalised Killing vector $K$ is related to the calibration forms for supersymmetric branes.
The general calibrations for $\mathrm{M}5$ and $\mathrm{M}2$ branes are given by the~\eqref{genkcal} and~\eqref{genkcalM2}. 
With an appropriate choice of the $\mathrm{AdS}_5$ gamma matrices (see~\cref{app:mtheoryconv}), the 11-dimensional Killing vector $\mathcal{K}_M$ in~\eqref{11d1f} has only the following non-zero components,
\begin{subequations}
\begin{align}
& \mathcal{K}_0 = \bar{\psi} \rho_0 \psi \\
& \mathcal{K}_m = -\im \left(\bar{\chi}_1+\chi_2^T\right)\gamma_{m}\left(\chi_1 - \chi_2^* \right) = \xi_m \, ,
\end{align}
\end{subequations}
where we fixed the norms of the spinors to $\bar \chi \chi = 1$ and $(\bar{\psi}\psi) = \im/2$ and $\xi_m$ is the Reeb vector on $M_6$. Consistently, we also fixed the value of the angular parameter to $\a = \pi/4$ in~\eqref{chiredef}.

The specific expression of the calibration forms $\Phi_{\mathrm{M}5}$ and $\Phi_{\mathrm{M}2}$ in~\eqref{genkcal} and~\eqref{genkcalM2} depends on how many AdS directions are spanned by the world-volume of the branes.

Consider an $\mathrm{M}5$ wrapping a $5$-cycle in $M_6$. 
We choose again the static gauge for the brane embedding and we set to zero the world-volume gauge field (so $H = A$). 
The the relevant components of $\Sigma$ and $\omega$ in~\eqref{11d2f} and~\eqref{11d5f} are the internal ones,
	\begin{subequations}
		\begin{align}
			\omega_{m_1 m_2} &= \e^\Delta \, \bar{\chi}\gamma_7 \gamma_{m_1 m_2} \chi = \e^\Delta Y^\prime \, , \\
			\Sigma_{m_1 \ldots m_5} &= \e^\Delta \, \bar{\chi}\gamma_7 \gamma_{m_1 \ldots m_5} \chi = \e^\Delta Z \, , 
		\end{align}
	\end{subequations}
and the calibration form in~\eqref{genkcal} reads (recall that the pull-back of $\tilde A$ is zero),
	\beq
		\Phi_{\mathrm{M}5} = \e^{\Delta} Z -\e^{\Delta} A \wedge Y' + \tfrac{1}{2}\iota_{\xi} A\wedge A \, . 
	\eeq
Note that this is exactly the pull-back on the brane of the twisted generalised vector $K$ in~\eqref{eq:twisted_K_M}. 
A similar computation for an $\mathrm{M}2$-brane wrapping a $2$-cycle in $M_6$ gives
	\beq
		\Phi_{\mathrm{M}2} = \e^{\Delta} Y' - \iota_{\xi} A \, ,
	\eeq
which is again the pull-back on the $\mathrm{M}2$-brane of the twisted generalised vector $K$.
Using the $L_K$ conditions~\eqref{eq:m-theory_structure_eqs} and choosing a gauge for $A$ such that 
$\mathcal{L}_\xi A =0$, it is straightforward to check that $\Phi_{\mathrm{M}5}$ and $\Phi_{\mathrm{M}2}$ are closed.
Explicitly, for instance for $\mathrm{M}5$, one has,
	\beq
		\begin{split}
			\diff \Phi_{\mathrm{M}5} &= \diff (\e^{\Delta} Z ) - \diff (\e^{\Delta} A \wedge Y') + \tfrac{1}{2}\diff(\iota_{\xi} A\wedge A) \\
					&= \e^{\Delta} Y' \wedge F - F \wedge \e^\Delta Y' + \iota_\xi F \wedge A + \tfrac{1}{2}\diff(\iota_\xi A)\wedge A + \tfrac{1}{2}\iota_\xi A \wedge F \\
					&= \iota_\xi F \wedge A + \tfrac{1}{2} \mathcal{L}_\xi A \wedge A - \tfrac{1}{2}\iota_\xi F \wedge A - \tfrac{1}{2} A \wedge \iota_\xi F = 0\, .
		\end{split}
	\eeq
Analogously, one can verify that $\Phi_{\mathrm{M}2}$ is also closed, showing that the purely internal configuration of the membrane is supersymmetric.

The generalised vector $K$ is also related to the calibration forms for other types of brane probes. 
Here, as example, we briefly discuss the case of branes with one one leg in the external space-time, that is a string moving in AdS. 
We leave the study of other more general membrane configurations to future work.
The calibration forms for $\mathrm{M}2$ and $\mathrm{M}5$-branes of this kind are given by~\eqref{genkcal} and~\eqref{genkcalM2} in this case, take the following form
	\begin{align}
		& \Phi_{\mathrm{M}2} = \e^{2\Delta} \tilde{\zeta}_1 \\
		& \Phi_{\mathrm{M}5} = \e^{2\Delta} \star Y' + \e^{2\Delta} \tilde{\zeta}_1 \wedge A 
	\end{align}
where $Z = \star \tilde{\zeta}_1$. 
The two calibrations are components of the (poly)-form 
	\beq
	\label{eq:cal_M_q1}
		\Phi= \e^{2\Delta}\tilde{\zeta}_1 +\e^{2\Delta}\star_6 Y'\, + \e^{2\Delta} \tilde{\zeta}_1 \wedge A\, .
	\eeq
which is the Hodge dual of the vector structure~\eqref{eq:twisted_K_M}. 
We want now to study its closure and its relation to the integrability conditions.
In this case, the closure follows from the moment map condition $\m_3 \equiv 0$, rather than from the $L_{K}$ condition. 
In~\cite{Ashmore:2016qvs}, it is shown that this condition requires
\beq
	\diff \bigl( \e^{2\Delta} \tilde{\zeta}_1\bigr) = 0\, ,
\eeq
so that this form calibrates a $\mathrm{M}2$-brane. 
Again, combining the two conditions, we get
\beq
	\begin{array}{lcr}
		\diff \lb \e^{3\Delta} \sin\Theta \rb = 2 m \e^{2\Delta} \tilde{\zeta}_1 & \mbox{and} & \diff \lb \e^{3\Delta} V\rb = \e^{3\Delta} \sin \Theta F + 2m \e^{2\Delta} \star Y'\, .
	 \end{array}
\eeq
From the vanishing of $\m_3$ in~\cite{Ashmore:2016qvs}, it is easy to verify that the form $\e^{2\Delta} \star_6 Y' + \e^{2\Delta}\tilde{\zeta}_1 \wedge A$ is closed (for non-vanishing $m$).
%
%
%%%%%%
%
%%%%%%%%%%%%%%%%%%%%%% M theory on AdS_4 %%%%%%%%%%%%%%%%%%%%%%%%%%%%%%%%%%%%%%%%%%%
\subsection{\texorpdfstring{Calibrations in $\mathrm{AdS}_4 \times M_7$}{Calibrations on AdS4 x M7}}
\label{M-thAdS4}
In this section, we discuss M-theory calibrations on $\mathrm{AdS}_4$ backgrounds. 
Again, we first review the exceptional generalised geometry~\cite{Ashmore:2016qvs} and then relate it to generalised calibrations.
Conventions for the spinor bilinears and the supersymmetry equations for the internal forms can be found in~\cite{Gabella:2012rc} and the relevant ones for this work are collected in appendix~\ref{app:mtheoryconv}. 

The background metric has the following form 
	\beq
	\label{eq:metric_mtheory_ads4}
		\diff s^2 = \e^{2\Delta} \diff s^2_{\mathrm{AdS}_4} + \diff s^2_{M_7} \, .
	\eeq
We set the inverse $\mathrm{AdS}_4$ radius to $m=2$. In addition, there is a non trivial four-form flux 
	\beq
		G = m \mathrm{vol}_4 + F \, ,
	\eeq
where $F = \diff A$ is the flux component on $M_7$ and it satisfies the following Bianchi identity and equations of motion 
	\beq
		\begin{array}{ccccc}
			\diff F = 0\, , & & & &\diff( \e^{2\Delta} \star_7 F ) = - m F \, , 
	\end{array}
	\eeq
with $\star_7$ the Hodge star on $M_7$. 
We will also need its dual $\tilde{F} = \diff \tilde{A} - \tfrac{1}{2} A \wedge F$.

The $11$-dimensional gamma matrices split as 
	\beq
	\label{eq:dec_gammas_m7}
		\begin{array}{ccc}
			\Gamma_{\m} = \e^{\Delta} \rho_{\m} \otimes \id & \mbox{and} & \Gamma_m = \e^{\Delta} \rho_5 \otimes \g_m\, ,
		\end{array}
	\eeq
with $\{\rho_{\m}, \rho_{\n}\}=2 g_{\m \n}$ and $\{\g_m, \g_n\}= g_{mn}$. 
The matrix $\rho_5 = \im \rho_{0123}$ is the chirality operator in four dimensions, and $\g_{1 \ldots 7}=\im \id$.
For further details about Clifford algebra conventions we refer to the~\cref{app:conv}.

The spinor ansatz preserving eight supercharges reads~\cite{Gabella:2012rc, Gabella:2011sg}
	\beq
	\label{eq:spinor_m7}
		\begin{split}
			\varepsilon &= \sum_{i=1,2} \psi_i \otimes \e^{\Delta/2} \chi_i + \psi^c_i \otimes \e^{\Delta/2} \chi_i^c\\
					&= \e^{\Delta/2} \psi_+ \otimes \chi_- + \e^{\Delta/2} \psi_- \otimes \chi_+ + \mathrm{c.c.} 
		\end{split}
	\eeq
where $\chi_{\pm} \coloneqq \tfrac{1}{\sqrt{2}} \lb \chi_1 \pm \im \chi_2\rb$ and $\psi_{\pm} \coloneqq \tfrac{1}{\sqrt{2}} \lb \psi_1 \pm \psi_2\rb$. 
In addition, we take the $\mathrm{AdS}_4$ spinors $\psi_i$ to have positive chirality, \emph{i.e.} $\rho_5 \psi_i = \psi_i$.\\

Combining the supersymmetry conditions and equations of motion for the fluxes one can express the internal fluxes in terms of spinor bilinears by~\cite{Gabella:2012rc},
	\begin{align}
		F &= \dfrac{3 m}{\tilde{f}}\ \diff (\e^{6 \Delta} \Im( \bar{\chi}_+^c \gamma_{(3)} \chi_-))\, , \\
		\tilde{F} &= - \tilde{f}\ \mathrm{vol}_7 \, . 
	\end{align}

The features of the solutions depend on the \emph{electric} charge $m$. 
When $m=0$ the solutions correspond to near horizon geometries of $\mathrm{M}5$-branes wrapped on internal cycles (no $\mathrm{M}2$ charge). 
The geometries with $m\neq0$ correspond to the presence of a non-vanishing $\mathrm{M}2$ charge. 
For $m \neq 0$ the internal manifolds always admit a canonical contact structure, as shown in~\cite{Gabella:2012rc}. \\

%%%
%%
%
%\subsubsection{Exceptional Sasaki-Einstein structure}
%\label{ESEAdS4}
%
	The generalised geometry relevant for backgrounds of this kind is discussed in~\cite{Ashmore:2016qvs}. 
	The $HV$ structure is given by a generalised vector $X$ in the fundamental of $E_{7(7)}$ and a triplet $J_a$ in the adjoint representation. 
	The untwisted vector reads
		\beq
			\tilde{X}= \xi + \e^{3\Delta} Y + \e^{6\Delta} Z - \im \e^{9\Delta} \tau\, ,
		\eeq	
where the forms are bliinears in the internal background spinors 
		\beq
		\label{bilrel}
			\begin{array}{lcccccr}
				\sigma = \im \bar{\chi}_+^c \gamma_{(1)} \chi_- \, , &\phantom{and}& Y = \im \bar{\chi}_+^c \gamma_{(2)} \chi_- \, ,&\phantom{and}& Z=\star_7 Y \, ,& \phantom{and}& \tau = \sigma \otimes \mathrm{vol}_7\, ,
			\end{array}
		\eeq
and $\xi$ is the vector dual to the one-form $\sigma$. 
Notice that the vector structure has the same form in both cases of a Sasaki-Einsten internal manifold and of a generic flux background~\cite{Ashmore:2016qvs}. 
Indeed the seven-dimensional manifolds giving $\mathcal{N}=2$ supersymmetry always admit a local $SU(2)$ structure. 
Moreover, the Killing vector constructed by spinor bilinears in~\eqref{bilrel} (or equivalently its dual one-form $\sigma$) defines a contact structure.
This allows us to write the $M_7$ metric as a Reeb foliation, analogously to the case of a Sasaki-Einstein manifold~\cite{Gabella:2012rc}. 
As a consequence, the volume form can be written making use of the contact structure,
		\beq
			\dfrac{1}{3!} \sigma \wedge \diff \sigma \wedge \diff \sigma \wedge \diff \sigma = \left(\dfrac{3m^2}{\tilde{f}}\right)^3 \e^{9\Delta} \mathrm{vol}_7 = 2\left(\dfrac{3m^2}{\tilde{f}}\right)^3 \sqrt{q(K)}\, ,
		\eeq
where $q(K)$ is the $E_{7(7)}$ invariant and $K$ is the real part of the twisted vector structure $X$
		\beq
		\label{Kads4}
			K = \xi - \frac{1}{2} \sigma \wedge \omega \wedge \omega + \iota_\xi \tilde{A}\, .
		\eeq
As already mentioned in~\cref{ExStr}, supersymmetry implies that $X$ is a generalised vector and its vector part, $\xi$, is a Killing vector. 
Through AdS/CFT, $\xi$ is the dual of the R-symmetry of the conformal $\mathcal{N}=2$ gauge theory in three dimensions. 
Then, as discussed in~\cref{sec:def_exceptional_geometry}, the generalised Lie derivative along $X$ must reduce to $\mathcal{L}_\xi$, which implies the vanishing of the tensor $\tilde{R}$, or more explicitly
		\beq
		\label{M7Rvan}
			\begin{array}{l}
				\diff (\e^{3\Delta}Y)=\iota_{\xi} F\, , \\
				\diff (\e^{6\Delta}Z)= \iota_{\xi} \tilde{F} -\e^{3\Delta} Y \wedge F\, .
			\end{array}
		\eeq
As expected, these reproduce part of the supersymmetry equations in~\cite{Gabella:2012rc}.

One can choose the gamma matrices and spinors in such a way that the Killing vector $\mathcal{K}$ has components~\cite{Gabella:2012rc} 
	\beq
		\begin{split}
			\mathcal{K}_0 & = \sum_i \bar{\psi}_i\rho_0 \psi_i\, , \\
			\mathcal{K}_m & = - \tfrac{\im}{2} \e^{2\Delta}\, \bar{\chi}^c_+ \gamma_m \chi_- \, .
		\end{split}
	\eeq

As in the previous section, the form of the generalised calibrations for $\mathrm{M}5$ and $\mathrm{M}2$ branes,~\eqref{genkcal} and~\eqref{genkcalM2}, depends on the direction spanned by the branes. 
Again, we considered first an $\mathrm{M}5$ wrapping a 5-cycle in $M_7$, with a zero world-volume gauge field ($H = A$) and in the static gauge. 

In this case, the relevant components of the forms~\eqref{11d2f} and~\eqref{11d5f} are 
	\begin{equation*}
		\begin{aligned}
			& \omega = \tfrac{\im}{2} \e^{3\Delta} \bar{\chi}_{+}^c \gamma_{(2)} \chi_{-} = \e^{3\Delta} Y \, ,\\[1mm]
			& \Sigma = -\e^{6\Delta}(\bar{\chi}_+ \gamma_{(5)} \chi_+^c + \bar{\chi}_-^c \gamma_{(5)} \chi_- ) = \e^{6\Delta} Z\, ,
		\end{aligned}
	\end{equation*}
and the calibration $\Phi_{\mathrm{M}5}$ gives
	\beq
		\Phi_{\mathrm{M}5} = (\e^{6\Delta} Z + A \wedge \e^{3\Delta} Y + \tfrac{1}{2}\iota_\xi A \wedge A + \iota_{\xi}\tilde{A})\, .
	\eeq	
One can also add an $\mathrm{M}2$ completely arranged along the internal directions. 
The corresponding calibration form is given by,
	\beq
		\Phi_{\mathrm{M}2} = (\e^{3\Delta} Y + \iota_{\xi} A )\, ,
	\eeq
which together with $\Phi_{\mathrm{M}5}$ gives rise to a poly-form,
	\beq
		\Phi = (\e^{3\Delta} Y + \iota_{\xi} A ) + (\e^{6\Delta} Z + A \wedge \e^{3\Delta} Y + \tfrac{1}{2}\iota_\xi A \wedge A + \iota_{\xi}\tilde{A})\, ,
	\eeq
and this, again, corresponds to the vector structure. 
The closure of $\Phi$, follows from supersymmetry%
		\footnote{%
			Note that $L_K$ conditions imply that $\Phi_{\mathrm{M}5}$ and $\Phi_{\mathrm{M}2}$ are separately closed.%
			}%
			, more precisely, from the $L_K$ conditions~\eqref{M7Rvan},
		\beq
			\begin{split}
				\diff \Phi &= \diff (\e^{3\Delta} Y + \iota_{\xi} A ) + \diff (\e^{6\Delta} Z + A \wedge \e^{3\Delta} Y + \tfrac{1}{2}\iota_\xi A \wedge A + \iota_{\xi}\tilde{A}) \\[1mm]
						& = \iota_\xi F + \diff (\iota_\xi A) + \iota_\xi \tilde{F} - \e^{3\Delta} Y \wedge F + F \wedge \e^{3\Delta} Y - A \wedge \iota_\xi F \\
						& \phantom{=} + \tfrac{1}{2} \diff (\iota_\xi A) \wedge A + \tfrac{1}{2} \iota_\xi A \wedge F + \diff (\iota_\xi \tilde{A}) \\[1mm]
						&= \mathcal{L}_\xi A + \mathcal{L}_\xi \tilde{A} + \tfrac{1}{2} \mathcal{L}_\xi A \wedge F + \tfrac{1}{2} A \wedge \iota_\xi F - A \wedge \iota_\xi F + \tfrac{1}{2}\iota_\xi A \wedge F = 0\, ,
			\end{split}	
		\eeq
where in the last line we made a gauge choice, such that,
	\begin{align}
	\label{gaugeM}
		& & & \mathcal{L}_\xi A= 0\, , & & \mathcal{L}_\xi \tilde{A}= 0\, . & 
	\end{align}

One can consider not only branes wrapping internal cycles. 
For instance, for an $\mathrm{M}5$-brane spanning two spatial directions, we can show that -- also in this case -- the related calibration form comes from the vector $K$. 
The relevant form is given by
		\beq
			\Phi = (\e^{3\Delta} Y + \iota_{\xi} A) \, .
		\eeq
The closure of this form comes directly from~\eqref{M7Rvan}.

It is also easy to see that, in this case, branes with one leg aligned with an external space direction 
are not supersymmetric, as indeed already discussed in~\cite{SanchezLoureda:2005ap}.

For instance in the case of an $\mathrm{M}2$ wrapping an internal cycle and with one external leg, the candidate calibration form is proportional to $\sigma$ in~\eqref{bilrel}, which is not closed -- \emph{i.e.} $\diff \sigma \sim \omega$. 
This condition, in the language of Exceptional generalised geometry, is a part of $ L_K J_a = \epsilon_{abc}  \lambda_b J_c $. 
It is interesting to note that, on the other hand, this configuration is supersymmetric in the case of a Minkowski background, since the analogous condition reads $L_K J_a = 0$,~\cite{Ashmore:2015joa}.

To conclude the analysis, let us focus on space-filling brane configurations. 
This case corresponds to the $J_a$ components of the Exceptional Sasaki-Einstein structure. 
For instance, for an $\mathrm{M}5$-brane we find
		\beq
		\label{q3cal}
			\Phi = -\e^{4\Delta}{V}_-\, ,
		\eeq
where $V_{-}$ is the two-form defined from spinor bilinears as follows,
		\beq
		\label{Vbil}
			V_{\pm} \coloneqq \dfrac{1}{2\im} \lb \bar{\chi}_+ \g_{(2)}\chi_+ \pm \bar{\chi}_- \g_{(2)}\chi_-\rb\, ,
		\eeq
	which gives the $TM \otimes T^*M$-component of $J_a$ by raising one index. 
	In particular, in the limit of a Sasaki-Einstein manifold (the only one for which the expression of $J_a$ is given explicitly in~\cite{Ashmore:2016qvs}), the calibration form~\eqref{q3cal} corresponds to $J_3$, and we have good reasons to trust this result also for the cases where generic fluxes are turned on. 
	We leave the complete discussion of these cases for future work.

%%%%%%%%%%%%%%%%%%%%%%%%%
%%%%%%%%%%%%%%%%%%%%%%%%
%%%%%%%%%%%%%%%%%%%%%%%%%

%%%%%%%%
%%
%
%%%%%%%%%%%%%%%%%%%%%%%%%%%%%%%%%%%%%%%%%%%%%%%%%%%%%%%%%%%%%%%%%
%%======================================================================
%%======================================================================
%%++++++++++++++++++++++++++++++++++++++++++++++++++++++++++++++++++++++
%%++++++++++++++++++++++++++++++++++++++++++++++++++++++++++++++++++++++
%%======================================================================
%%======================================================================
%%===========================Type IIB section===========================================
%%======================================================================
%%++++++++++++++++++++++++++++++++++++++++++++++++++++++++++++++++++++++
%%++++++++++++++++++++++++++++++++++++++++++++++++++++++++++++++++++++++
%%======================================================================
%%======================================================================
%
\section{Supersymmetric branes in type IIB}
\label{sec:IIBcal}
In this section, we want to discuss the analogous conditions to have supersymmetric extended objects in a type IIB supergravity $\mathrm{AdS}$ background and point out their connections to Exceptional Sasaki-Einstein structures defining such backgrounds.

We are interested in backgrounds with non trivial fluxes. The NS three-form is $H = \diff B$ and the RR fields are 
	\beq
		\begin{aligned}
	 	F_1=\diff C_0\, , & & & &
	 	F_3=\diff C_2\, , & & & &
	 	F_5=\diff C_4 - \frac{1}{2} H \wedge C_2 + \frac{1}{2} F_3 \wedge B \, .
		\end{aligned}
	\eeq
The field strengths $F$ satisfy the duality condition
	\beq
	\label{fdual}
		F_p = (-1)^{\left[\tfrac{p}{2}\right]} \star F_{10-p}\, , 
	\eeq
while the S-duality of type IIB is reflected in the fact that $B$ with $C_2$ form an $SL(2)$ doublet. It could be useful to define a complexified version of the three-form flux~\cite{Schwarz:1983qr},
	\beq
		G = F_3 + \im H_3\, .
	\eeq
The Bianchi identities are written as
	\begin{align}
		& &	\diff F_5= \frac{1}{8}\, \Im\, G\wedge G^*\, ,
			& & 
		\diff G= 0\, . & & 
	\end{align}

The generalised calibrations for backgrounds with non-trivial NS-NS flux have been constructed in~\cite{Martucci:2011dn} (see also~\cite{Cascales:2004qp, HPS04} for an equivalent derivation in terms of the supersymmetry algebra). 
The two Majorana-Weyl supersymmetry parameters $\varepsilon_1$ and $\varepsilon_2$ can be used to construct the following bilinears~\cite{BJRT99},
	\begin{align}
		\label{K10}
		\mathcal{K} &= \dfrac{1}{2} ( \bar{\varepsilon}_1 \Gamma_M \varepsilon_1 + \bar{\varepsilon}_2 \Gamma_M \varepsilon_2)\ \diff x^M\, , \\
		\label{omega10}
		\omega &= \dfrac{1}{2} ( \bar{\varepsilon}_1 \Gamma_M \varepsilon_1 - \bar{\varepsilon}_2 \Gamma_M \varepsilon_2)\ \diff x^M\, , \\
		\label{psiIIB}
		\Psi &= \sum_{k=0}^2 \dfrac{1}{(2 k+ 1)!} \bar{\varepsilon}_1 \Gamma_{M_1 \ldots M_{2k +1} } \varepsilon_2 \ \diff x^{M_1} \wedge \ldots \wedge \diff x^{M_{2k +1}} \, . 
	\end{align}
Using the Killing spinor equations for type IIB, one can show that the vector $\hat{\mathcal{K}}$ dual to the one form $\mathcal{K}$ is a Killing vector~\cite{Tomasiello:2011eb},
	\beq
		\begin{array}{ccc}
			\mathcal{L}_{\hat{\mathcal{K}}} g = 0\, ,& \phantom{and}& \mathcal{L}_{\hat{\mathcal{K}}} F = 0 \, . 
		\end{array}
	\eeq
Notice that also the spinor bilinears~\eqref{K10}--\eqref{psiIIB} are invariant under the transformation generated by~$\hat{K}$
	\beq
		\begin{array}{ccc}
			\mathcal{L}_{\hat{\mathcal{K}}} \omega = 0\, , & \phantom{and} & \mathcal{L}_{\hat{\mathcal{K}}} \Psi = 0 \, .
		\end{array}
	\eeq

As discussed in~\cite{Martucci:2011dn}, we may write the $\kappa$-symmetry condition to have a supersymmetric $\mathrm{D}p$-brane 
	\beq
		\hat{\Gamma}_{\mathrm{D}p}\ \varepsilon_2 =\varepsilon_1\, ,
	\eeq
where, the $\kappa$-symmetry operator is defined as~\cite{BT97, Marolf:2003vf} 
	\beq
	\label{eq:brane_kappa}
		\hat{\Gamma}_{\mathrm{D}p}=\frac{1}{\sqrt{-\det\lb P[G]+\cf \rb}} \sum_{2l+s=p+1}\frac{\epsilon^{\a_1 \ldots \a_{2l}\b_1 \ldots \b_s}}{l!s!2^l} \cf_{\a_1\a_2} \ldots \cf_{\a_{2l-1}\a_{2l}} \Gamma_{\b_1 \ldots \b_s}\, ,
	\eeq
and $P[\bullet]$ denotes the pullback to the $(p+1)$-dimensional brane world-volume and $\cf= F + P[B]$, with $B$ the NS two-form and $F$ the world-volume gauge field-strength.

The energy of the brane (the charge associated to the transformation generated by $\hat{\mathcal{K}}$) is
	\beq
		E = - \int_{\mathcal{S}}\ \diff^p \sigma\ \hat{P}_M \hat{\mathcal{K}}^M \, ,
	\eeq
where $\mathcal{S}$ is the brane world-space and $\hat{P} = \tfrac{\partial L_{\mathrm{D}p}}{\partial (\partial_\tau X^M)}$ takes the form
	\beq
		\hat{P}_M = - \mu_{\mathrm{D}p} e^{- \phi} \sqrt{- \det \mathcal{M}} (\mathcal{M}^{-1})^{( \a \tau)} B_{MN} \partial_\a X^N 
			+ \frac{\mu_{\mathrm{D}p}}{p!} \epsilon^{\tau \a_1 \ldots \a_p} \left[ \iota_M (C \wedge e^\mathcal{F}) \right]_{\a_1 \ldots a_p} \, ,
	\eeq
where we denoted $\mathcal{M} = P[g] + \mathcal{F}$. 
Note that again we are in the temporal gauge in adapted coordinates, such that the world-volume of the brane is $\mathbb{R}\times \mathcal{S}$.
One has the usual BPS bound,
	\beq
		E \geq E_{BPS}\, ,
	\eeq
with
	\beq
		\label{BPScal}
			\begin{split}
				E_{BPS} =\mu_{\mathrm{D}p} \int_\mathcal{S} \diff^p \sigma\, &P\left[e^{- \phi} \Psi - \iota_{\hat{\mathcal{K}}} C - \omega \wedge C \right] \wedge \e^{\mathcal{F}} \\
					&+ \mu_{\mathrm{D}p}\int_{\mathcal{S}}\diff^p \sigma\, P\left[\omega-\iota_{\hat{\mathcal{K}}}B\right] \wedge \left.\left(C \wedge \e^{\mathcal{F}}\right)\right\vert_{p-1} \, .
			\end{split}
	\eeq
Thus, one can read the generalised calibration form from the last expression,
	\beq
	\label{BPScal2}
		\Phi_{\mathrm{D}p} = e^{- \phi} \Psi - \iota_{\hat{\mathcal{K}}} C - \omega \wedge C \wedge \e^{\mathcal{F}} + \omega-\iota_{\hat{\mathcal{K}}}B \wedge \left.\left(C \wedge \e^{\mathcal{F}}\right)\right\vert_{p-1}\, .
	\eeq
One can show~\cite{Martucci:2011dn, Evslin:2007ti} that this is a topological quantity.
In addition, one can also show that this form is closed, making use of potential configurations preserving the symmetry generated by $\hat{\mathcal{K}}$, \emph{i.e.}
	\begin{align}
	\label{vanpot}
		&&\mathcal{L}_{\hat{\mathcal{K}}} B = 0\, , & &\mathcal{L}_{\hat{\mathcal{K}}} C = 0\, ,&&
	\end{align}
analogously to the gauge choice~\eqref{gaugeM} in the previous section for M-theory. 
As a final observation, we would like to point out that the same conclusions about calibration forms can be obtained by supertranslation algebra, as done for example in~\cite{Cascales:2004qp, Gutowski:1999tu,HPS04}. \\

Let us now focus on type IIB compactifications to $\mathrm{AdS}_5$-backgrounds. 
As for the discussion of supersymmetric extended objects in M-theory above, we now apply the supersymmetry conditions and the aforementioned approach to branes in type IIB string theory on 
	\beq
	\label{eq:metric_IIB_ads5}
		\diff s^2= \e^{2\Delta} \diff s^2_{\mathrm{AdS}_5} + \diff s^2_{M_5}\, ,
	\eeq
and relate the calibration forms to the geometric description by the vector and hypermultiplet structures. 
The exceptional geometry of this setup is discussed in~\cite{Ashmore:2016qvs,Grana_Ntokos}, based on the geometric description in~\cite{Gauntlett:2005ww}. 
For $\mathcal{N}=2$ backgrounds of the form~\eqref{eq:metric_IIB_ads5} with generic fluxes, the internal manifold $M_5$ admits a (local) identity structure~\cite{Gauntlett:2005ww,GGPSW09, GGPSW09_02}. 

%
%
%
%\subsection{Exceptional Sasaki-Einstein structure}
%\label{sect:IIB_ESE}
%
The two ten-dimensional Majorana-Weyl spinors of the same chirality which describe a IIB background of the form~\eqref{eq:metric_IIB_ads5} can be decomposed as in~\cite{Grana_Ntokos}%
		\footnote{%
		We follow the conventions given in the appendix of~\cite{Grana_Ntokos}.
		We collect them in~\cref{sect:IIB_notation}.
		},
	\beq
	\label{eq:splitting_IIB_ads5}
		\varepsilon_i= \psi \otimes \chi_i \otimes u + \psi^c \otimes \chi_i^c \otimes u\, . 
	\eeq
Here $\psi$ denotes the external $\mathrm{Spin(4,1)}$ spinor, $\chi_i$ are the internal $\mathrm{Spin}(5)$ spinors and $u$ a two-component spinor. 
It might be convenient to define the complex spinors $\zeta_1 = \chi_1 + \im \chi_2$ and $\zeta_2^c = \chi_1^c + \im \chi_2^c$.
As for the previous cases, one can construct the relevant bilinears defining a local identity structure on the internal manifold~\cite{Gauntlett:2005ww}. 
One introduces the vectors
\beq
\label{IIBbil}
	\begin{aligned}
 		K_0^m &:= \bar{\zeta}_1^c\gamma^m\zeta_2\, , \\
 		K^m_3 &:= \bar{\zeta}_2\gamma^m\zeta_1 \, ,\\
 		K^m_4 &:= \tfrac{1}{2}\left(\bar{\zeta}_1\gamma^m\zeta_1 - \bar{\zeta}_2\gamma^m\zeta_2\right)\, , \\
 		K^m_5 &:= \tfrac{1}{2}\left(\bar{\zeta}_1\gamma^m\zeta_1 + \bar{\zeta}_2\gamma^m\zeta_2\right)\, ,
\end{aligned}
\eeq
that are not all linearly independent. 
Relations between these forms comes from supersymmetry, as shown in~\cite{Gauntlett:2005ww}. 
Then, one can use the following scalars to parametrise the norms of the spinors
\beq
\label{scalarbilIIB}
	\begin{aligned}
 		A &:= \tfrac{1}{2}\left(\bar{\zeta}_1\zeta_1 + \bar{\zeta}_2\zeta_2\right)\, , \\
		A\sin\Theta &:= \tfrac{1}{2}\left(\bar{\zeta}_1\zeta_1 - \bar{\zeta}_2\zeta_2\right)\, , \\
		S &:= \bar{\zeta}^c_2\zeta_1\, , \\
		Z &:= \bar{\zeta}_2\zeta_1\, .
	\end{aligned}
\eeq
Finally, one considers the two-forms
\beq
\label{IIB2formsbil}
	\begin{aligned}
		U_{mn} &:= -\tfrac{\im}{2}\left(\bar{\zeta}_1\gamma_{mn}\zeta_1 + \bar{\zeta}_2\gamma_{mn}\zeta_2\right)\, , \\
		V_{mn} &:= -\tfrac{\im}{2}\left(\bar{\zeta}_1\gamma_{mn}\zeta_1 - \bar{\zeta}_2\gamma_{mn}\zeta_2\right)\, , \\
		W_{mn} &:= -\bar{\zeta}_2\gamma_{mn}\zeta_1\, .
	\end{aligned}
\eeq

The HV structure for these backgrounds can be found in \cite{Ashmore:2016qvs, Grana_Ntokos}. 
The untwisted generalised vector structure 
$K \in \Gamma(\tilde{E})$ in \eqref{genvecs} is given in terms of the identity structure above by 
	\beq
	\label{eq:K_tilde_IIB}
		\begin{aligned}
			\tilde{K}= \tilde{\xi} + \tilde{\l}^i +\tilde{\rho} + \tilde{\sigma}^i = K_5^{\sharp} +\e^{2\Delta - \frac{\phi}{2} } \begin{pmatrix} \Re K_3 \\ \Im K_3 \end{pmatrix} - \e^{4\Delta - \phi} \star V \, ,
		\end{aligned}
	\eeq
where $\phi$ is the dilaton and $\Delta$ the warp factor.
Notice that for these backgrounds the five-forms vanish $\sigma^i =0$. 
The twisted vector structure is obtained by acting on $\tilde K$ with the adjoint element as in appendix E of~\cite{Ashmore:2015joa}
	\beq
	\label{eq:IIB_twisted_explicit}
		%\begin{split}
	K = \xi + \l^i + \rho \, ,
	 %+ \eta^i +\left[\tilde{\l}^i + \iota_{\xi}B^i\right]
	 		%+ \left[\tilde{\rho} + \epsilon_{kl} \tilde{\l}^k \wedge B^l +\frac{1}{2} \epsilon_{kl}\lb \iota_{\xi} B^k\rb \wedge B ^l+ \iota_{\xi}C \right]\\
	 		%&+ \left[\tilde{\rho} \wedge B^i - C \wedge \tilde{\l}^i +\frac{1}{2} \lb \epsilon_{kl}\tilde{\l}^k \wedge B^l +\iota_{\xi} C\rb \wedge B^i 
	 		% -\frac{1}{2} C \wedge \lb \iota_{\xi}B^i\rb + \frac{1}{6} \epsilon_{kl} \lb \iota_{\xi} B^k \rb \wedge B^l \wedge B^i \right].
		%\end{split}
	\eeq
where the twisted quantities are~\cite{Ashmore:2016qvs,Ashmore:2015joa}
	\begin{subequations}
	\label{IIbtwist}
		\begin{align}
			\xi &= \tilde{\xi}\, , \label{IIbtwistv} \\[1mm]
			\l^i & = \tilde{\l}^i + \iota_{\xi}B^i\, , \label{IIbtwist1} \\[1mm]
			\rho & = \tilde{\rho} + \iota_\xi C + \epsilon_{ij}\tilde{\lambda}^i \wedge B^j + \tfrac{1}{2}\epsilon_{ij}\lb \iota_\xi B^i \rb\wedge B^j\, . \label{IIbtwist3}%\\[1mm]
			%\begin{split}
			%\label{IIbtwist5} 
				%\eta^i & = \tilde{\rho} \wedge B^i - C \wedge \tilde{\l}^i +\tfrac{1}{2} \lb \epsilon_{kl}\tilde{\l}^k \wedge B^l +\iota_{\xi} C\rb \wedge B^i + \\
				% & \phantom{=} -\tfrac{1}{2} C \wedge \lb \iota_{\xi}B^i\rb + \tfrac{1}{6} \epsilon_{kl} \lb \iota_{\xi} B^k \rb \wedge B^l \wedge B^i\, . 
			%\end{split}
		\end{align}
	\end{subequations}
and we defined $B^1 = B$, $B^2 = C_2$, $C = C_4$, $F^1= H$, $F^2 = F_3$ and $F=F_5$.

As already discussed the condition that the generalised Lie derivative $L$ along the Reeb vector $K$ has to reduce to the conventional one, $\mathcal{L}_{\xi}$ implies some differential equations on the elements of the vector structure that reproduce some of the supersymmetry conditions on the identity structure derived in~\cite{Gauntlett:2005ww},
	\beq
	\label{eq:IIB_ESE_structure_eqs}
		\begin{split}
			\diff \tilde{\l}^i&= \iota_{\xi} F^i\, ,\\
			\diff \tilde{\rho} &= \iota_{\xi} F +\epsilon_{ij} \tilde{\l}^i \wedge F^j\, .
		\end{split}
	\eeq
%
%======================================================================
%======================================================================
%++++++++++++++++++++++++++++++++++++++++++++++++++++++++++++++++++++++
%++++++++++++++++++++++++++++++++++++++++++++++++++++++++++++++++++++++
%======================================================================
%======================================================================
%
%\subsection{\texorpdfstring{$\kappa$-symmetry and brane calibrations}{k-symmetry and brane calibrations}}
%%
%\label{sec:kappa_symmetry_IIB}

Analogously to the M-theory case, we want now to express the calibration conditions for a D$p$ probe in these backgrounds in terms of the generalised structure and check that their closure is implied by differential conditions on Exceptional Sasaki-Einstein structures. To this purpose we have to specialise the calibrations \eqref{K10}, \eqref{omega10} and \eqref{psiIIB} to the various brane configurations.
The $\mathrm{AdS}_5$ geometry and, in particular the products of external spinors, is the same as in the previous section. Thus, in our conventions, the Killing vector $\mathcal{K}$ has the following components,
	\beq
		\begin{split}
			\mathcal{K}_0 &= \e^{-2\Delta} \bar{\psi} \rho_0 \psi \otimes A \, , \\
			\mathcal{K}_m & = \tfrac{1}{2}\left(\bar{\zeta}_1\gamma^m\zeta_1 + \bar{\zeta}_2\gamma^m\zeta_2\right) = \xi_m \, ,
		\end{split}
	\eeq
where $\xi$ is the Reeb vector. 
We also fix the norm of the internal spinors such that $A=1$.

As in the previous sections, we focus on the cases of point-like AdS particles and space-filling branes where the calibrations are related to the generalised vector $K$. 
Consider first a D$1$ wrapping an internal one-cycle. 
The relevant terms in~\eqref{BPScal2} are 
	\beq
		\begin{aligned}
			&\omega =- \e^{2\Delta-\phi/2} \tfrac{1}{2} \left(\bar{\zeta}_2 \gamma_m \zeta_1 + \zeta_2^T\gamma_m \zeta_1^* \right) = - \e^{2\Delta-\phi/2} \Re K_3\, , \\
			&\Psi =- \e^{2\Delta+\phi/2}\tfrac{1}{2\im} \left(\bar{\zeta}_2 \gamma_m \zeta_1 - \zeta_2^T\gamma_m \zeta_1^* \right) = - \e^{2\Delta+\phi/2} \Im K_3\, , \\
			&\iota_{\hat{\mathcal{K}}} B = \iota_\xi B^1 =: \iota_\xi B\, , \\
			&\iota_{\hat{\mathcal{K}}} C = \iota_\xi B^2\, .
		\end{aligned}
	\eeq
and it is immediate to see that the calibration form is given by the generalised vector $K$
	\beq
		\Phi_{\mathrm{D}1} = -\tilde{\lambda}^2 -\iota_\xi B^2 = -\e^{2\Delta-\phi/2} \Im K_3 -\iota_\xi B^2\, .
	\eeq
Using equation~\eqref{eq:IIB_ESE_structure_eqs} one can show that $\Phi_{\mathrm{D}1}$ is closed. 
Using again the properties of the $\mathrm{AdS}_5$ spinors, it is easy to show that a space-filling D$5$-brane is also calibrated by the same form,
	\beq
		\Phi_{\mathrm{D}5} = ( -\tilde{\lambda}^2 -\iota_\xi B^2) \otimes \star 1 =( -\tilde{\lambda}^2 -\iota_\xi B^2) \otimes \mathrm{vol}_{\mathrm{AdS}_5} \, .
	\eeq

Similarly, one can find the calibration form for a D$3$-brane wrapping purely internal cycles,
	\beq
	\label{D3cal}
		\Phi_{\mathrm{D}3}= \tilde{\rho} +\iota_\xi C +\epsilon_{ij}\tilde{\lambda}^i \wedge B^j + \tfrac{1}{2}\epsilon_{ij}\iota_\xi B^i \wedge B^j\, .
	\eeq
Its closure follows from the $L_K$ conditions under the gauge choice%
		\footnote{%
		We also need to use $\tilde{\rho} \propto \star V$.%
		} that the potentials are invariant under $K$.
Again, this form provides also the calibration for a space-filling D$7$-brane.

In the particular case where the only non-trivial background flux is the five-form, the generalised Sasaki-Einstein structure reduces to the standard one and the internal manifold is Sasaki-Einstein.
In this case the spinor ansatz~\eqref{eq:splitting_IIB_ads5} simplifies since the two internal spinors are proportional to each other, \emph{i.e.} $\chi_2=\im \chi_1$, and, consequently, the one-form part vanishes.
The twisteed vector~\eqref{eq:IIB_twisted_explicit} simplifies to
	\beq
		K = \xi - \sigma \wedge \omega + \iota_{\xi}C\, , 
	\eeq 
and the (untwisted) hypermultiplet structure is~\cite{Ashmore:2016qvs}
	\begin{align}
	\label{Jstruct}
		\tilde{J}_+ &= \tfrac{1}{2}\kappa u^i \Omega - \tfrac{\im}{2} \kappa u^i \Omega^{\sharp}\, ,\\[1mm]
		\tilde{J}_3 &= \tfrac{1}{2}\kappa I + \tfrac{1}{2} \kappa \hat{\tau}^{i}_{\phantom{i}j} + \tfrac{1}{8}\kappa \Omega^{\sharp}\wedge \bar{\Omega}^{\sharp} - \tfrac{1}{8}\kappa \Omega \wedge \bar{\Omega} \, , 
	\end{align}
where $u^i = (-i, 1)^i$ and $I$, $\omega$ and $\Omega$ are the complex structure, the symplectic and the holomorphic two-forms on the K\"ahler-Einstein basis of $M_5$. 

Note that in the vector structure, the three-form is $\sigma \wedge \omega$ and by the structure equation~\eqref{eq:IIB_ESE_structure_eqs} one immediately sees that adding the potential part~$\iota_{\xi}C$ yields a closed form (up to a gauge choice),
	\beq
		\diff \lb \tilde{\rho}+\iota_{\xi}C \rb= \iota_{\xi} \diff C + \mathcal{L}_{\xi} C - \iota_{\xi} \diff C = \mathcal{L}_{\xi} C = 0\, .
	\eeq

Since in this case, the form of the hypermultiplet structure is simple, we can also study calibrations that are not associated to the vector $K$. 
We will do it in the simplest Sasaki-Einstein background, namely $\mathrm{AdS}_5 \times S^5$, where $S^5$ is the five-dimensional sphere. 
The background is given by
	\beq
	\label{ads5bkgrd}
		\begin{split}
			\diff s^2 &= \dfrac{R^2}{r^2} \diff r^2 + \dfrac{r^2}{R^2} \eta_{\mu\nu} \diff x^\mu \diff x^\nu + \diff s^2(S^5)\, , \\[1mm]
			C_4 &= \left(\dfrac{r^4}{R^4} - 1\right) \diff x^0 \wedge \ldots \wedge \diff x^3\, .
		\end{split}
	\eeq
while all other fluxes, dilaton and warp factors vanish.
The $S^5$ can be written as a $U(1)$ fibration over $\mathbb{CP}^2$,
	\beq
		\diff s^2(S^5) = \diff \Sigma_{4}^2 + \sigma \otimes \sigma\, ,
	\eeq
where $\diff \Sigma_{4}^2$ is the \emph{Fubini-Study metric} over $\mathbb{CP}^2$. The form $\sigma$ is given by $\sigma = \diff \psi + A$, where $A$ is a connection such that $\mathcal{F} = \diff A = 2 \omega$, and $\psi$ is the periodic coordinate on the circle $U(1)$ with period $6\pi$.

Explicitly, the sphere $S^5$ takes the form~\cite{Gauntlett:2004yd}
	\beq
		\begin{split}
			\diff s^2(S^5) &= \diff \a^2 + \dfrac{1}{4} \sin^2 \a (\diff \theta^2 + \sin^2 \theta \diff \phi^2) + \dfrac{1}{4} \cos^2 \a \sin^2\a (\diff \beta + \cos\theta \diff \phi)^2 \\
				 & \phantom{=} + \dfrac{1}{9}\left[\diff \psi - \dfrac{3}{2}\sin^2\a (\diff \beta + \cos\theta \diff \phi)\right]^2\, ,
		\end{split}
	\eeq
with $\psi \in [0,6\pi]$, $\beta \in [0,4\pi]$, $\a \in [0,\pi/2]$, $\theta \in [0,\pi]$ and $\phi \in [0,2\pi]$.
In these coordinates the holomorphic form has the following expression,
	\beq
		\begin{split}
		\Omega &= \frac{1}{2} e^{\im (\psi - \beta) } \sin \sigma  \diff \theta \wedge d\sigma + \frac{1}{4} \im e^{\im( \psi - \beta) } \sin^2 \sigma  \cos \sigma  \diff \beta \wedge \diff \theta -\frac{1}{2} \im e^{\im (\psi - \beta) } \sin \theta \sin \sigma  \diff \sigma \wedge \diff \phi \\ 
				& \phantom{=} - \frac{1}{4} e^{\im (\psi - \beta) } \sin \theta \sin^2 \sigma \cos \sigma \diff \beta \wedge \diff \phi - \frac{1}{4} \im e^{\im (\psi - \beta) } \cos \theta  \sin^2 \sigma  \cos \sigma \diff \theta \wedge \diff \phi\, .
		\end{split}
	\eeq

First, consider a D$5$-brane spanning the directions $0,1,2, r$ in $\mathrm{AdS}_5$. 
The world-volume of the brane is $\mathrm{AdS}_4 \times S^2$, where $S^2$ is the sphere parametrized by the angles $(\theta, \phi)$. 
Then the expression~\eqref{BPScal2} reduces to 
	\beq
		\begin{split}
			\Phi_{\mathrm{D}5} &= -\ \diff x^0 \wedge \diff x^1 \wedge \diff x^2 \wedge \diff x^4 \wedge \tfrac{(1-\im)}{2} \e^{4\Delta}\Omega \\
				& = -\ \diff x^0 \wedge \diff x^1 \wedge \diff x^2 \wedge \diff r \wedge \mathrm{vol}_{S^2}\, ,
		\end{split}
	\eeq
and we see that it corresponds to the two-form part of $\tilde{J}_+$ in~\eqref{Jstruct}.
Modulo choice of coordinates,%
		\footnote{%
		Here $\diff x^4 \propto \sin \theta\ \diff r + \diff \sigma + \diff \beta$.%
		} it agrees with the analogous form in~\cite{Cascales:2004qp}.

We can also consider a D$3$-brane probe spanning the directions $0,1,r $ of $AdS_5$. 
The world-volume is now $\mathrm{AdS}_3 \times S^1$ and the calibration is given by the Hodge dual of the $4$-form part of $\tilde{J_3}$,
	\beq
		\Phi_{\mathrm{D}3} = \diff x^0 \wedge \diff x^1 \wedge \diff r \wedge \tfrac{1}{8} \e^{4\Delta} \star( \Omega \wedge \bar{\Omega})\, .
	\eeq
One can prove the closure of this form by the $L_K J$ relations. In particular,
	\beq
		\diff (\e^{4\Delta} \star ( \Omega \wedge \bar{\Omega})) = - m\ \iota_\xi \mathrm{vol}_5 = 0\, ,
	\eeq
where the first equality comes from the conditions to have a vanishing $\tilde{R}$-tensor~\cite{Ashmore:2015joa,Ashmore:2016qvs}. 
In other words, it is the rewriting of equation~\eqref{eq:IIB_ESE_structure_eqs} in the Sasaki-Einstein case.

\section*{Acknowledgements}
The authors thank Michela Petrini for proposing this topic and for many helpful discussions. We would like to thank also Eirik Eik Svanes and Matthieu Sarkis for interesting discussions.
O.d.F. is supported by the	ILP LABEX (under reference ANR-10-LABX-63) through French state funds managed by the ANR within the ``Investissements d'Avenir programme" under reference ANR-11-IDEX-0004-02. J.C.G. thanks the group LPTHE for the hospitality during his stay in Paris. His work was supported by the Deutsche Forschungsgemeinschaft (DFG, Germany) under the grant LE 838/13 and by the Research Training Group RTG 1463 ``Analysis, Geometry and String Theory'' (DFG).
%%======================================================================
%%======================================================================
%%++++++++++++++++++++++++++++++++++++++++++++++++++++++++++++++++++++++
%%+++++++++++++++++++++++++++++++Appendix++++++++++++++++++++++++++++++++
%%======================================================================
%======================================================================
\appendix
%======================================================================
%======================================================================
%++++++++++++++++++++++++++++++++++++++++++++++++++++++++++++++++++++++
%++++++++++++++++++++++++++++++++++++++++++++++++++++++++++++++++++++++
%======================================================================
%======================================================================
\section{Conventions for spinors and gamma matrices}
\label{app:conv}
%======================================================================
%======================================================================
%++++++++++++++++++++++++++++++++++++++++++++++++++++++++++++++++++++++
%++++++++++++++++++++++++++++++++++++++++++++++++++++++++++++++++++++++
%======================================================================
%======================================================================
In this appendix we collect the conventions for spinors and gamma matrices 
that are relevant for this paper. 
%%%%%%%%%%%%%%%%%%%%%%
%%%%%%%%%%%%%%%%%%%%%%%%

\subsection{\texorpdfstring{Type IIB on $\mathrm{AdS}_5 \times M_5$}{Type IIB on AdS5 x M5}}
\label{sect:IIB_notation}

We follow the conventions in~\cite{Grana_Ntokos}. The ten-dimensional metric is
\beq
	\diff s^2 = \e^{2A}\diff s^2_{\mathrm{AdS}_5} + \diff s^2_{M_5} \, , 
\eeq
and the ten-dimensional gamma matrices $\Gamma^M$ are chosen as 
\beq
	\begin{array}{lcc}
		\Gamma^{\m} = e^{-A} \rho^{\m} \otimes \id_4 \otimes \sigma^3\, ,& \phantom{\mbox{with}} & \mu=0, \ldots 4\, , \\
		\Gamma^{m+ 4} = \id_4 \otimes \g^m \otimes \sigma^1\, , & \phantom{\mbox{with}} & m=1, \ldots 5 \, ,
	\end{array}
\eeq
where $\rho^\m$ and $\gamma^m$ generate $\mathrm{Cliff}(1,4)$ and $\mathrm{Cliff}(5)$ respectively, satisfying 
\beq
	\begin{array}{ccc}
		\{\rho^{\m},\rho^{\n}\}= 2 g^{\m\n}\, , & \phantom{\mbox{with}}& \{\g^{m},\g^{n}\}= 2 g^{mn} \, , 
	\end{array}
\eeq
and $\eta^{\mu \nu} = \mathrm{diag}(-1, 1,1,1,1)$. We also have 
\beq
	\begin{array}{ccc}
		\rho^{01 \ldots 4}=-\im \id\, , & \phantom{\mbox{with}}& \g^{1 \ldots 5}=\id\, .
	\end{array}
\eeq
We choose the $\mathrm{Cliff}(1,4)$ and $\mathrm{Cliff}(5)$ intertwiners as
\beq
\label{Achoice}
	\begin{array}{lcl}
		A_{1,4} = \rho_0 & \quad & C_{1,4} = D_{1,4} A_{1,4}\, , \\
		A_5 = 1 & \quad & C_5 = D_5 \, ,
	\end{array}
\eeq
where $C = - C^T$ in any dimension, so that 
\beq
\label{IIbintertw}
	\begin{array}{lcl}
		\rho^{\mu \, \dagger} = - A_{1,4}\rho^{\m} A_{1,4}^{-1} & \quad \qquad \qquad & \g^{m \, \dagger} = \g^m\, , \\ 
		\rho^{\m \, T}= C_{1,4}\rho^{\m}C_{1,4}^{-1} & \quad \qquad \qquad & \g^{m \, T}= C_5 \g^m C_5^{-1}\, , \\ 
		\rho^{\m *}= - D_{1,4} \rho^{\m} D_{1,4}^{-1} & \quad \qquad \qquad & \g^{m *} = C_5 \g^m C_5^{-1} \, .
	\end{array}
\eeq

An explicit choice of a basis for the $\mathrm{Cliff}(1,4)$ gamma matrices is
\beq
\label{ads5gamma}
	\begin{array}{ccccc}
		\rho^0=\im \sigma^2 \otimes \sigma^0\, , & \rho^i = \sigma^1 \otimes \sigma^i\, , & \rho^4=-\sigma^3 \otimes \sigma^0\, , &\phantom{\mbox{and}} & i=1,2,3 \, ,
	\end{array}
\eeq
while for the $\mathrm{Cliff}(5)$ gamma matrices we take 
\beq
	\begin{array}{ccccc}
		\g^1=\sigma^1 \otimes \sigma^0\, , & \g^2=\sigma^2 \otimes \sigma^0\, , & \g^3=\sigma^3 \otimes \sigma^1\, , & \g^4=\sigma^3 \otimes \sigma^2\, , & \g^5=-\sigma^3 \otimes \sigma^3\, , 
	\end{array}
\eeq
with intertwiners 
\begin{equation*}
	\begin{array}{ccc}
		A_{1,4}=\rho^0\, , & C_{1,4}=\rho^0 \rho^2\, ,& C_5=\sigma^1 \otimes \sigma^2\, .
	\end{array}
\end{equation*}

With these choices for the gamma matrices the ten-dimensional chiral gamma decomposes as 
\beq
	\Gamma_{11} = \Gamma_{0, \ldots 9} = \id_4 \otimes \id_4 \otimes \sigma^2 \, . 
\eeq
The ten-dimensional supersymmetry parameters are Majorana-Weyl spinors of negative chirality $\Gamma_{11} \varepsilon_i = - \varepsilon_i$ ($i=1,2$) and decompose as
\beq
	\varepsilon_i = \psi \otimes \chi_i \otimes u + \psi^c \otimes \chi_i^c \otimes u\, ,
\eeq
where $\psi$ is an external $\mathrm{Spin}(4,1)$ spinor, $\chi_i$ are internal $\mathrm{Spin}(5)$ spinors and $u$ a two-component spinor satisfying
\beq
	\sigma^2 u = - u \qquad \qquad u^* = \sigma^1 u \, . 
\eeq

Charge conjugation of the external and internal spinors is defined as
\begin{align}
	\psi^c= D_{1,4} \psi^* && \chi^c= C_5 \chi^* \, . 
\end{align}
One can easily check that, with the above choices, 
\beq
\label{eq:IIB_spinor_cc_properties}
	\begin{array}{lcl}
		\psi^{cc}=-\psi\, , & \quad \quad & \lb \rho^{\m_1} \ldots \rho^{\m_k}\psi\rb^c= \lb -1\rb^k \rho^{\m_1} \ldots \rho^{\m_k} \psi^c \, , \\
		\chi^{cc}=-\chi\, , & \quad \quad & \lb \g^{m_1} \ldots \g^{m_k} \chi\rb^c= \g^{m_1} \ldots \g^{m_k} \chi^c\, .
	\end{array}
\eeq

For $5$-dimensional internal spinors, from the properties listed above, one can derive
\beq
	\lb \overline{\chi^c} \g_{m_1 \ldots m_r} \phi^c\rb = \lb \overline{\chi}\g_{m_1 \ldots m_r} \phi\rb^*\, ,
\eeq
and
\beq
	\lb \overline{\chi^c}\g_{m_1 \ldots m_r}\phi\rb = -\lb \overline{\chi} \g_{m_1 \ldots m_r} \phi^c\rb^*\, .
\eeq

Similarly we can derive some useful identities for the internal spinors.
Let us consider the expression,
\beq
	\lb \overline{\psi^c}\rho_{\m_1 \ldots \m_q} \psi^c\rb = \lb -1\rb^{q+1} \lb \overline{\psi} \rho_{\m_1} \ldots \rho_{\m_q} \psi\rb^*\, .
\eeq
where, as always, $\overline{\psi}=\psi^{\+}\rho_0$. Next, we obtain
\beq
	\lb \overline{\psi}\rho_{\m_1} \ldots \rho_{\m_q}\psi\rb^* = - \lb -1\rb^{\frac{q(q+1)}{2}} \lb \overline{\psi} \rho_{\m_1} \ldots \rho_{\m_q}\psi\rb\, .
\eeq
Combining these two equations yields
\beq
	\lb \overline{\psi^c}\rho_{\m_1 \ldots \m_q} \psi^c\rb = -\lb -1\rb^{\frac{(q+1)(q+2)}{2}} \lb \overline{\psi} \rho_{\m_1} \ldots \rho_{\m_q}\psi\rb\, .
\eeq
For the other combinations, one obtains
\beq
	\lb \overline{\psi^c}\rho_{\m_1 \ldots \m_q}\psi\rb^* = - \lb -1\rb^{q+1} \lb \overline{\psi} \rho_{\m_1} \ldots \rho_{\m_q} \psi^c\rb\, ,
\eeq
and 
\beq
	\lb \overline{\psi^c} \rho_{\m_1 \ldots \m_q}\psi\rb^* = -\lb -1\rb^{\frac{q(q+1)}{2}} \lb \overline{\psi} \rho_{\m_1 \ldots \m_q} \psi^c\rb\, . 
\eeq
Again combining the last two relations gives
\beq
	 \lb \overline{\psi}\rho_{\m_1 \ldots \m_q} \psi^c\rb = \lb -1\rb^{\frac{(q+1)(q+2)}{2}}\lb \overline{\psi}\rho_{\m_1 \ldots \m_q} \psi^c\rb\, ,
\eeq
so that these terms vanish for $q=0,1,4$.

%%%%%%%%%%%%%%%%%%%%%
%========M-theory section========%
%%%%%%%%%%%%%%%%%%%%%
\subsection{M-theory}
\label{app:mtheoryconv}
We follow again the conventions in~\cite{Grana_Ntokos} for the metric ansatz
\beq
	\diff s^2 = \e^{2\Delta}\diff s^2_{\mathrm{AdS}} + \diff s^2_{M} \, .
\eeq
We consider two M-theory setups: $\mathrm{AdS}_4$ compactifications with a $7$-dimensional internal manifold $M_7$
and $\mathrm{AdS}_5$ ones on a $6$-dimensional internal manifold $M_6$.
The eleven-dimensional gamma matrices are $\hat{\Gamma}^M$, $M=0,\ldots, 10$, satisfying the Clifford algebra $\mathrm{Cliff}(1,10)$ relations, 
\beq
	\{\hat{\Gamma}^A , \hat{\Gamma}^B \} = 2 \eta^{AB}\, .
\eeq
They will decompose as in~\eqref{eq:dec_gammas_m6} and~\eqref{eq:dec_gammas_m7}, and for convenience, we report them here,
\beq
\label{11gammadec}
	\begin{array}{lccr}
		\hat{\Gamma}^\mu = \e^{-\Delta}\,\rho^\mu \otimes \Gamma_7\, , & \hat{\Gamma}^{m+4} = \mathds{1}_4 \otimes \Gamma^m\, & \phantom{\mbox{for}} &\mbox{for}~\mathrm{AdS}_5 \times M_6\, , \\[2mm]
		\hat{\Gamma}^\mu = \e^{-\Delta}\,\rho^\mu \otimes \mathds{1}_8\, , & \hat{\Gamma}^{m+3} =\e^{-\Delta} \rho_5 \otimes \Gamma^m\, &\phantom{\mbox{for}} &\mbox{for}~\mathrm{AdS}_4 \times M_7\, .
	\end{array}
\eeq 
In the expressions above we denoted the internal gamma matrices with the same symbol for both cases, with $\Gamma_7$ the chiral operator in $6$ dimensions, defined below, and with $\rho_5$ the external $\mathrm{Cliff}(1,4)$ chiral operator.
%%%%%
%%%%%%%% 
%%%%%
\subsubsection{\texorpdfstring{M-theory on $\mathrm{AdS}_5 \times M_6$}{M-theory on AdS5 x M6}}
Here we give the conventions for M-theory solutions of the form $\mathrm{AdS}_5 \times M_6$. The external part is the same as the type IIB compactification.
So we refer to~\cref{sect:IIB_notation}. For the internal part we take as reference~\cite{VanProeyen:1999ni} and we write all the gamma matrices as tensor products of Pauli matrices
\beq
\label{gamma6}
	\begin{aligned}
		\Gamma^1 &= \sigma_1 \otimes \mathds{1} \otimes \sigma_{1}\, , \\[1mm]
		\Gamma^2 &= \sigma_1 \otimes \mathds{1} \otimes \sigma_{2}\, , \\[1mm]
		\Gamma^3 &= \sigma_1 \otimes \sigma_{1} \otimes \sigma_{3}\, , \\[1mm]
		\Gamma^4 &= \sigma_2 \otimes \sigma_{1} \otimes \sigma_{3}\, , \\[1mm]
		\Gamma^5 &= -\sigma_1 \otimes \sigma_{3} \otimes \sigma_{3}\, , \\[1mm]
		\Gamma^6 &= \sigma_3 \otimes \mathds{1} \otimes \mathds{1}\, . 
	\end{aligned}
\eeq
In even dimensions we can define a chiral operator
\beq
\label{chiralgamma}
	\Gamma_7 = - \im \Gamma^1 \ldots \Gamma^6\, .
\eeq
The chiral operator $\Gamma_7$ squares to the identity and satisfies the following relations with the other gamma matrices
\begin{align*}
	\left\{\Gamma_7\, , \Gamma_{m} \right\} &= 0\, ,\\
	\left[\Gamma_7\, , \Gamma_{mn} \right] &= 0\, .
\end{align*}
By induction, these properties can be extended to any odd/even rank element of the Clifford algebra.

The intertwiners of $\mathrm{Cliff(6)}$ can be written as follows
\begin{align*}
\Gamma_m^T & = C_6^{-1} \Gamma_m C_6 \, , \\
\Gamma_m^* & = D_6^{-1} \Gamma_m D_6 \, , \\
\Gamma_m^{\dagger} & = A_6 \Gamma_m A^{-1}_6 \, .
\end{align*}
and, for our conventions, 
\begin{equation*}
	\begin{array}{lcr}
		A_6 = 1 \, ,& & D_6 = C_6\, .
	\end{array}
\end{equation*}

%======================================
%%%%%%%%%%%%%%%%%%%%%%%%%%
%======================================
%%%%%%%%%%%%%%%%%%%%%%%%%%
\subsubsection{\texorpdfstring{M-theory on $\mathrm{AdS}_4 \times M_7$}{M-theory on on AdS4 x M7}}
Here we give the conventions which are relevant for the $\mathrm{AdS}_4$ solutions of M-theory. We made the choice of having compatible conventions with the previous section, such that one can embed all the relations above in the following ones.

The Clifford algebra $\mathrm{Cliff}(7)$ and its generators are constructed by the same set of gamma matrices~\eqref{gamma6} of $\mathrm{Cliff}(6)$ 
plus the chiral gamma $\Gamma_7$ in~\eqref{chiralgamma}.

The $\mathrm{Cliff}(7)$ intertwiners are written as
\begin{align*}
	\Gamma_m^T &= C_7^{-1} \Gamma_m C_{7}\, , \\[1mm]
	\Gamma_m^\dagger &= A_7 \Gamma_m A_7^{-1}\, , \\[1mm]
	\Gamma_m^* &= D_7^{-1} \Gamma_m D_7\, .
\end{align*}
Numerically the matrices $A_7$, $C_7$, $D_7$ are the same as $A_6$, $C_6$, $D_6$.

The four-dimensional gamma matrices on $\mathrm{AdS}_4$ satisfy 
\beq
	\{\rho_a , \rho_b \} = 2 \eta_{ab}\mathds{1} \, ,
\eeq
where $a, b$ are frame indices. Hence it holds $\eta^{ab}e^\mu_a \otimes e^\nu_b = g^{\mu\nu}$, where $g^{\mu\nu}$ is the $\mathrm{AdS}_4$ inverse metric. 
In terms of flat frame indices, we choose a basis for explicit calculations for $\mathrm{Cliff}(1,3)$,
\beq
\begin{array}{lrcc}
	\rho^0=\im \sigma^2 \otimes \sigma^0\, , & \rho^i = \sigma^1 \otimes \sigma^i\, , &\phantom{\mbox{and}} & i=1,2,3 \, . 
\end{array}
\eeq
As for the internal part we have chosen the basis above such that we can embed it into~\eqref{ads5gamma}.
The intertwiners can be written as in~\eqref{Achoice},
\beq
	\begin{array}{ccc}
		\rho^{\mu\dagger} = -A_{1,3} \rho^\mu A_{1,3}^{-1}\, , &\phantom{\mbox{for}} &A_{1,3} = \rho_0\, , \\[1mm]
		\rho^{\mu T} = C_{1,3} \rho^\mu C_{1,3}^{-1}\, , &\phantom{\mbox{for}} & C_{1,3} = D_{1,3} A_{1,3} \, , \\[1mm]
		\rho^{\mu *} = -D_{1,3} \rho^\mu D_{1,3}^{-1}\, . & \phantom{\mbox{for}} & 
	\end{array}
\eeq
%======================================================================
%======================================================================
%++++++++++++++++++++++++++++++++++++++++++++++++++++++++++++++++++++++
%++++++++++++++++++++++++++++++++++++++++++++++++++++++++++++++++++++++
%======================================================================
%======================================================================

\bibliographystyle{utphys}
\bibliography{Bibliography}
\nocite{*}

\end{document}